\documentclass[%
 superscriptaddress,
twocolumn,
nofootinbib,
longbibliography,
 amsmath,amssymb,
 aps,
pra,
]{revtex4-2}

\usepackage{color}
\usepackage{graphicx}% Include figure files
\usepackage{bm}% bold math
\usepackage{subfigure}
\usepackage{hyperref}
\usepackage{mathrsfs}
\usepackage{xcolor}
\usepackage{csquotes}
\usepackage{cleveref}

\usepackage{xparse}
\usepackage{xstring}

\hypersetup{
	colorlinks = true,
	linkcolor  = black,
	urlcolor   = blue,
	citecolor  = green!90!black,
	breaklinks = true
}

\newcommand{\commText}[2]{\bf Note by #1: #2}
\NewDocumentCommand{\comment}{O{anonymous}m}{
    \IfEqCase{#1}{%
        {TF}{\textcolor{green!60!black}{\commText{#1}{#2}}}%
        {MK}{\textcolor{teal}{\commText{#1}{#2}}}%
        {FM}{\textcolor{blue}{\commText{#1}{#2}}}%
        {JL}{\textcolor{red}{\commText{#1}{#2}}}%
    }%
    [\textcolor{orange}{\commText{#1}{#2}}]
}

\begin{document}

\title{Artificial Intelligence and Machine Learning for Quantum Technologies}

\author{Mario Krenn}
\email{ML4qtech@mpl.mpg.de}
\affiliation{Max Planck Institute for the Science of Light, Erlangen, Germany.}
\author{Jonas Landgraf}
\affiliation{Max Planck Institute for the Science of Light, Erlangen, Germany.}
\affiliation{Department of Physics, Friedrich-Alexander Universität Erlangen-Nürnberg, Germany.}
\author{Thomas Foesel}
\affiliation{Max Planck Institute for the Science of Light, Erlangen, Germany.}
\affiliation{Department of Physics, Friedrich-Alexander Universität Erlangen-Nürnberg, Germany.}
\author{Florian Marquardt}
\affiliation{Max Planck Institute for the Science of Light, Erlangen, Germany.}
\affiliation{Department of Physics, Friedrich-Alexander Universität Erlangen-Nürnberg, Germany.}

\date{\today}

\begin{abstract}
In recent years, the dramatic progress in machine learning has begun to impact many areas of science and technology significantly. In the present perspective article, we explore how quantum technologies are benefiting from this revolution. We showcase in illustrative examples how scientists in the past few years have started to use machine learning and more broadly methods of artificial intelligence to analyze quantum measurements, estimate the parameters of quantum devices, discover new quantum experimental setups, protocols, and feedback strategies, and generally improve aspects of quantum computing, quantum communication, and quantum simulation. We highlight open challenges and future possibilities and conclude with some speculative visions for the next decade. 
\end{abstract}

\maketitle
\tableofcontents

\section{Introduction}

\begin{figure*}
    \centering
    \includegraphics[width=\textwidth]{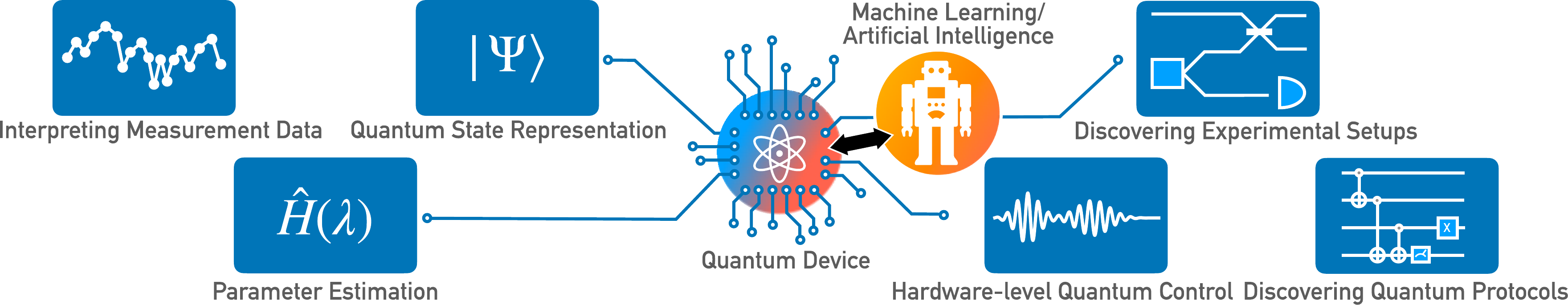}
    \caption{\label{fig:overview} Overview of tasks in the area of quantum technologies that machine learning and artificial intelligence can help solve better, as explained in this perspective article.}
\end{figure*}

The fields of machine learning \cite{jordan2015machine,lecun2015deep,goodfellow2016deep} and quantum technologies \cite{dowling2003quantum,flamini2018photonic,georgescu2014quantum,preskill2018quantum} have a lot in common: Both started out with an amazing vision of applications (in the 1950s and 1980s, respectively), went through a series of challenges, and are currently extremely hot research topics. Of these two, machine learning has firmly taken hold beyond academia and beyond prototypes, triggering a revolution in technological applications during the past decade. This perspective article will be concerned with shining a spotlight on how techniques of classical machine learning (ML) and artificial intelligence (AI) hold great promise for improving quantum technologies in the future. A wide range of ideas have been developed at this interface between the two fields during the past five years, see Fig.~\ref{fig:overview}. Whether one tries to understand a quantum state through measurements, discover optimal feedback strategies or quantum error correction protocols, or design new quantum experiments, machine learning can yield efficient solutions, optimized performance and, in the best cases, even new insights.

With the present review, we aim to take physicists with a background in quantum technologies on a tour of this rapidly growing area at the interface to classical machine learning. The readers are not expected to have a background in machine learning and will get a state-of-the-art view into how machine learning techniques are applied to quantum physics. We should state right away that in this perspective article we tried to achieve our goal by focusing in each application domain on some selected illustrative examples. Our selection is necessarily subjective. We thus make no claim as to providing a comprehensive list of the literature and apologize to anyone who misses some favorite work.

We hope that after seeing the examples discussed in our review, the reader will appreciate how useful machine learning techniques could be for quantum technologies. At the same time, we also want to make the reader aware of how crucial it is to choose the right AI approach for a problem. It is by no means necessary that the most advanced AI method is the most suited tool for a given task. Often, modern deep-learning methods can be significantly outperformed by rather simple methods if applied to the wrong task, as seen in other areas \cite{angelini2022cracking}. Here it is crucial to analyze the scope of the problem and decide on the best-suited algorithms.

Once the reader wants to understand how to apply these tools in practice, we recommend the very educational lecture notes introducing machine learning techniques with a view to their application to quantum devices, both brief \cite{marquardt2021machine}, and very extended \cite{dawid2022modern}. In addition, there are several reviews from recent years with a somewhat different focus than ours, e.g.~about machine learning applied to physics in general \cite{carleo2019machine} or machine learning for quantum many-body physics \cite{carrasquilla2020machine}. 

We also remark that there is the whole field of quantum machine learning, which tries to discover potential quantum advantages when implementing new learning algorithms on quantum platforms. This also includes variational quantum circuits, which are very promising in the context of noisy intermediate-scale quantum computers (NISQ). We do not cover these developments here, and we refer the interested reader to reviews for quantum machine learning  \cite{schuld2015introduction,biamonte2017quantum,dunjko2020non,lamata2020quantum} and variational quantum algorithms and NISQ devices \cite{cerezo2021variational,bharti2022noisy}. While we focus on quantum technology here, we want to mention that there is also considerable work on ML for foundational quantum science \cite{bharti2020machine}.

In the following section, we will briefly introduce the basics of neural networks and other machine learning techniques, aiming to set the stage for subsequent discussions. The bulk of our review is contained in section \ref{sec:applications}, where we discuss the various applications of machine learning to quantum technologies. In each case, we aim to remark on some of the challenges and potential future research directions. Finally, in the outlook, we speculate about how machine learning might have transformed quantum technologies in a dozen years from now. 

\section{Basic Techniques of Machine Learning and Artificial Intelligence}
\label{sec:techniques}

The purpose of this section is merely to provide a glimpse of the essential basics so that the subsequent discussion of applications becomes intelligible to the reader without prior exposure. Machine learning techniques have been around for several decades and involve many efficient approaches that predate the recent deep learning revolution, like "support vector machines" or "decision trees" \cite{book:Russel_norvig_AI_book}. However, the flexibility of neural networks has made them a popular general-purpose choice, so we focus on those in our brief introduction.

\subsection{Evolutionary Algorithms}

One set of algorithms that often been used in optimization tasks in the domain of artificial intelligence are evolutionary (genetic) algorithms \cite{whitley1994genetic,sivanandam2008genetic}. There, the idea is to deal with a set of candidate solutions, each of them described via a suitable vector. These solutions can be randomly changed ("mutated"), two solutions can be combined to form a new candidate ("crossover"), and finally only the best solutions can be kept for the next round of evolution ("selection"). Such methods can be surprisingly effective, and their applications to new problems is often straight forward. As we will describe in further sections, genetic algorithms have successfully been applied to quantum-technology-related tasks.

\subsection{Neural Networks: Structure}
% including conv., recurrent

\begin{figure}
    \centering
    \includegraphics[width=\columnwidth]{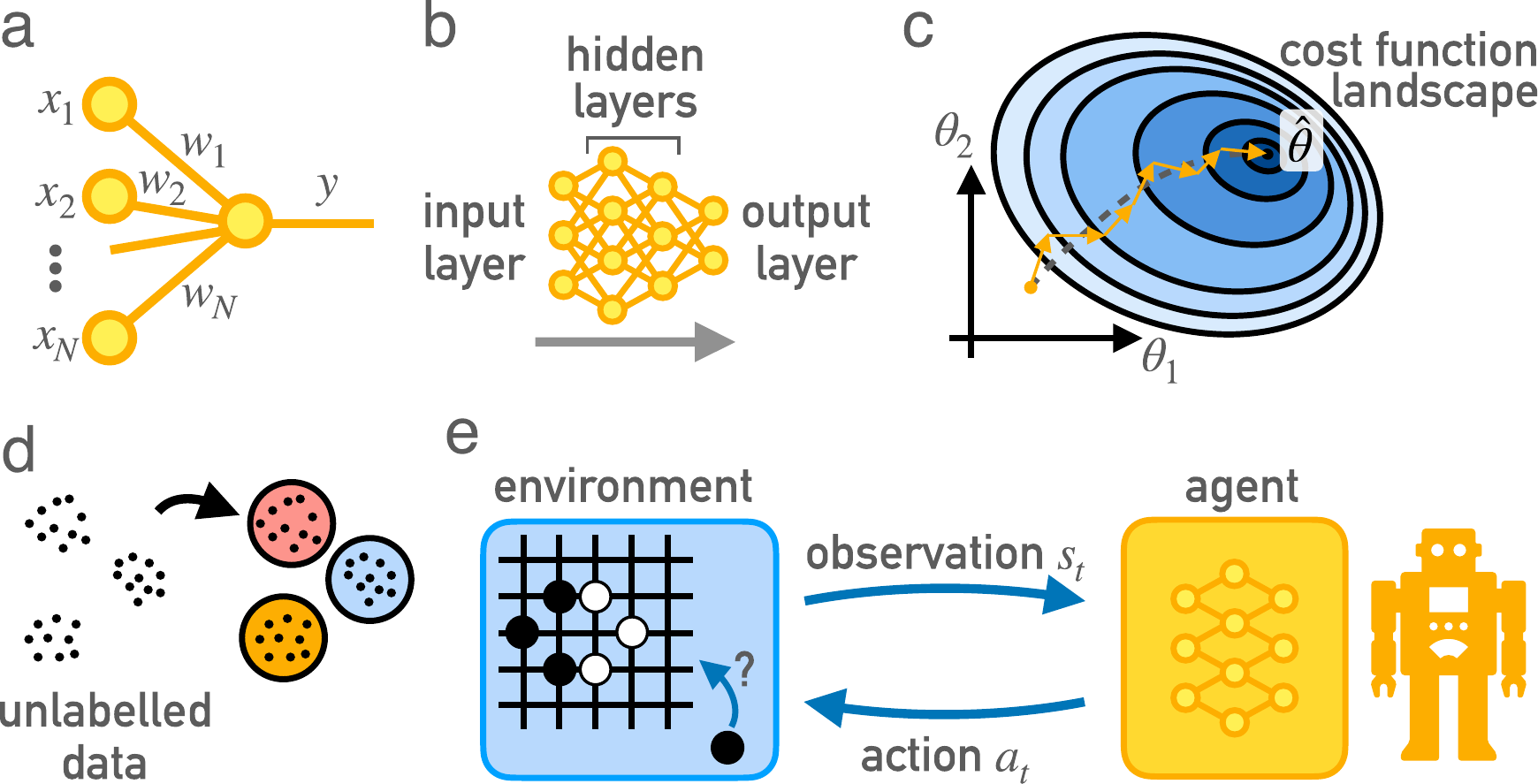}
    \caption{\label{fig:sketch_NN} Basics of Neural Networks and Machine Learning Techniques. (a) Operation of a single artificial neuron. (b) Structure of a neural network with dense layers. (c) Evolution of the network's parameters $\theta$ in the cost function landscape using stochastic gradient descent. For every step (orange arrows) the gradients of the averaged cost function with respect to $\theta$ (gray dashed line) are approximated by by the gradient averaged over a random batch. The parameters $\hat{\theta}$ minimize the averaged cost function. (d) Classification of unlabeled data. (e) Reinforcement learning problem modelled as Markov decision process.}
\end{figure}
The structure of artificial neural networks, which can be trained to approximate arbitrary functions, is loosely motivated by neurons in the human brain. Each neuron receives multiple inputs and generates an output signal which serves as input for other neurons. 
% The output function is in general highly nontrivial and reflects the human experience.
% FM says: I was unsure what 'the output function' here means. The previous sentence would suggest it's the activation function or y=f(z), but that doesn't make sense.
During learning, the strength of connections between neurons changes. 

More precisely, each artificial neuron receives $N$ inputs $x_j$ which are summed according to some weights $w_{j}$, representing connection strengths, see Fig.~\ref{fig:sketch_NN}a. After adding a bias (shift) $b$, a simple nonlinear "activation function" $f$ is applied to yield the output $y$:
\begin{subequations}
	\begin{align}
		y &= f(z)\\
		z&=\sum_{j=1}^N w_{j} x_j +b
	\end{align}
\end{subequations}
Without the nonlinear activation function between each layer, the whole neural network could be compressed into one single linear transformation, with very limited computational power. Popular activation functions are the \enquote{ReLU} ($f(z)=z$ for $z\geq0$ and $f(z)=0$ for $z\leq 0$) and the \enquote{sigmoid} function ($f(z)=1/(\mathrm{e}^{-z}+1)$), which represents a smoothed step-function.

Multiple connected neurons form a neural network as shown in Fig.~\ref{fig:sketch_NN}b. These networks are typically structured into several layers, where the output from each layer serves as input for the next layer. Overall, there is one input layer, one or more intermediate "hidden" layers, and an output layer. The input being fed into such a network could be, e.\,g.~pixel data from an image, and after successful training, the output neurons encode e.\,g.~the category of the image (cat vs.~dog). A network with only one hidden layer and suitably many neurons is already able to approximate arbitrary functions with arbitrary precision \cite{cybenko1989proof_nn_universality}. However, networks with more hidden layers (called \enquote{deep} neural networks) may be able to fulfill the same task with fewer neurons.

In the network in Fig.~\ref{fig:sketch_NN}b, all outputs of one layer serve as input for all neurons of the next layer. Such a structure is called a \textit{dense} layer, or "fully connected network." However, this is often not the best choice, especially when the input has certain symmetries. For example, an image classification task can have a translational symmetry since the category of an object shown in a picture is independent of the object's precise location. Convolutional neural networks (CNNs) use this translational invariance by convolving the input with a "kernel" that can be learned.

Time series have a temporal structure that can be utilized using so-called "recurrent" neural networks (RNNs). The key idea is that the RNN receives the information for each subsequent time step sequentially as input but also keeps some memory about previous inputs. One of the most advanced and commonly-used RNNs are so-called long short-term memory (LSTM) \cite{hochreiter1997long} networks.

\subsection{Neural Networks: Training}
\label{sec:learning_methods}

A network can be made to approximate any function by suitably choosing its weights and biases, often summarily denoted as a parameter vector $\theta$. To produce correct predictions with a neural network, its parameters have to be trained for the given problem. Supervised learning (SL) is a training method to learn a desired target function $f(x)$ from a dataset containing pairs  ("samples") of inputs and associated outputs, $(x, f(x))$. A neural network represents an approximate input-output relation $f_{\theta} (x)$. The deviation between the network's prediction and the correct output in the dataset is quantified by a "cost" (or "loss") function. For any given input $x$, this cost function can be calculated, $C(f_\theta(x), f(x))$, and eventually, it will be averaged over all samples in the data set. The trainable parameters $\theta$ of the NN $f_\theta(x)$ have to be chosen so as to minimize this (sample-averaged) cost function.

The choice of cost function depends on the problem. Typical fields of application for neural networks are regression and classification tasks. For regression, the target function $f(x)$ is a continuous function of the input $x$. Then, the so-called "mean-square error" is the canonical choice, which for $n$ output neurons reads:
\begin{equation}
    C_\mathrm{MSE}(x) = \sum_{i=1}^n \left(f_{\theta,i}(x) - f_i(x)\right)^2
\end{equation}
On the other hand, in classification tasks, the input should be assigned to certain predefined classes (e.g.~categories of images). This is solved by having each output neuron correspond to one of the $n$ classes. Each neuron value (or "activation") can then be interpreted as the probability that the input is assigned to the corresponding class. 
%To ensure, that the output represents a valid probability distribution, the so-called softmax function ($\exp z_i/\sum_{i=1}^n \exp z_i$) is chosen as activation function for the output layer. 
In that situation, the typical cost function is the so-called "cross-entropy," a means to compare probability distributions.

%\begin{equation}
%    C_\mathrm{CE} = - \frac{1}{n} \sum_{i=1}^n %f_i(x) \log f_{\theta,i}(x)
%\end{equation}
%It is minimal if the probability distribution $f_\theta(x)$ matches the target distribution $f(x)$. $f_i(x)$ is 1 if the dataset assigns the input $x$ to class $i$, else 0.

In any case, neural-network training relies on the minimization of the cost function using gradient descent. However, evaluating the cost function averaged over all samples of the data set is infeasible. Rather, in each update step, the cost function is averaged over a batch of randomly selected inputs. In the simplest version of the resulting "stochastic gradient descent" scheme (see Fig.~\ref{fig:sketch_NN}c), the parameters $\theta$ are updated in each step according to $\theta\rightarrow \theta - \eta g$ with the gradient $g$:
\begin{equation}
    g = \frac{\partial \langle C\rangle_\mathrm{batch}}{\partial \theta}
\end{equation}
$\langle \cdot \rangle_\mathrm{batch}$ denotes the average over the random batch, and $\eta$ is the "learning rate," controlling the update's size. 

A neural network has many parameters (even small networks usually have thousands of parameters, the largest published networks have hundreds of billions of parameters). Therefore, it is crucial that there exists a highly efficient approach to calculating the gradient: the so-called "backpropagation" scheme. As its name implies, this algorithm calculates the gradients layer by layer, starting from the output layer. Remarkably, it is computationally not more demanding than the original evaluation of the neural network (also called the forward propagation). A significant hardware advance for training neural networks are GPUs (graphical processor units) that are optimized to perform highly-efficient manipulations of large matrices. This efficiency is essential for the success of ML in many applications. 

For more details on neural network architectures, backpropagation, and gradient descent, the interested reader is referred to the many existing excellent introductions \cite{Nielsen-ML-book,goodfellow2016deep,marquardt2021machine,dawid2022modern}.

\subsection{Unsupervised Learning}
Unsupervised learning approaches can learn the structure of unlabelled data sets on their own, for example, as shown in Fig.~\ref{fig:sketch_NN}d. Typical fields of application are feature learning, where the machine is asked to find a compact representation of the data, and clustering, where the computer has to sort on its own samples into classes with similar properties. Generative models are also often attributed to the field of unsupervised learning. Their purpose is to stochastically create new samples that follow the same distribution as the previously observed data set (e.g.~images of the same type, though never seen before). The simplest approaches to feature learning still rely on the techniques of supervised learning (namely, "self-supervised" learning in so-called "autoencoders"). However, clustering and generative models are typically implemented using distinct techniques (see, e.g.~\cite{goodfellow2016deep}).

\subsection{Reinforcement Learning}

Some of the most important problems in artificial intelligence and machine learning can be seen as the attempt to find an optimal strategy to cope with a certain task, where the optimal strategy is unknown (thus, supervised learning cannot be applied). This is the domain of Reinforcement Learning (RL), which aims to discover the optimal action sequences in decision-making problems. Its power has been famously illustrated in board games like chess or Go \cite{silver2016alphago} or video games\cite{vinyals2019grandmaster}, in all of which RL is able to reach superhuman performance. In many applications, RL can find the optimal strategy without prior knowledge about the actual dynamics of the system. When that is the case, we speak of "model-free" RL. The goal in any RL task is encoded by choosing a suitable "reward," a quantity that measures how well the task has been solved.

The typical RL problem can be understood as a so-called "Markov Decision Process" (MDP), see Fig.~\ref{fig:sketch_NN}e. An MDP consists of an "agent" (the controller) and an "environment" (the world, or the system to be controlled), and both interact in multiple time steps. In each time step $t$, the environment's state $s_t$ is observed. Solely based on this observation, the agent decides on its next action $a_t$, which will change the environment's state. The agent's behavior is defined by the "policy" $\pi(a|s)$, which denotes the probability of choosing the action $a$ given the observation $s$. 
%Dependent on the application, this feedback loop is repeated until a maximum number of time steps is reached, the environment has entered a particular state, or the agent decided to terminate. 
For each action $a_t$, the agent receives a reward $r_t$. For example, in a game, this reward could be $+1/-1$ at the last time step, when the agent has won/lost the game, and otherwise 0 in all previous steps. RL aims to maximize the cumulative reward $R$ (also called "return")
\begin{equation}
    R = \sum_{t=1}^T r_t \, ,
\end{equation}
where $T$ is the total number of time steps.

Three major branches of RL algorithms exist: policy gradient, Q-learning, and actor-critic methods. For policy-gradient methods, the agent directly sets the policy $\pi_\theta(a|s)$. In deep RL, this agent is realized by a deep neural network with trainable parameters $\theta$. To find the optimal strategy, policy-gradient estimates the gradient of the average return $\langle R\rangle$ with respect to $\theta$. Here, $\langle \cdot\rangle$ denotes the average over all trajectories for the current policy. At first sight, it is unclear how to take the gradients through the reward without knowing the model. However, one can compute the gradients of the frequency of a certain reward via the policy function $\pi_\theta$. Thus, the gradient turns out to be:
\begin{equation}
    \label{equ:basic_policy_gradient}
    g = \frac{\partial \langle R\rangle}{\partial \theta} = \left\langle R \sum_{t=1}^T \frac{\partial}{\partial \theta} \ln \pi_\theta(a_t|s_t) \right\rangle
\end{equation}
It is important to note that $R$ depends on the full trajectory and the parameters $\theta$ (we suppressed these dependencies for brevity). Updates based on this equation will increase ("reinforce") the probabilities of actions that occur predominantly in high-reward trajectories. Another approach to finding the optimal policy is so-called "Q-learning." It employs the "Q function" $Q_\pi(s,a)$ that tries to estimate the quality of an action $a$: it is defined as the average expected future return starting from a state $s$ and action $a$ for a policy $\pi$. The optimal policy is then, by definition, to choose the action that maximizes $Q$ in a given state $s$. In practice, the Q function is initially unknown. During training, an approximation to this function is learned, often using a deep neural network to represent $Q$.

Finally, the third group of RL algorithms are the so-called "actor-critic" methods. These try to combine the benefits of both policy-gradient and Q-learning approaches.
The basic idea behind actor-critic methods is to estimate the expected reward given the current state, the so-called value function $V$. The success of any action is then measured by comparing the resulting reward against $V$. As might be expected, the value function is represented by a neural network (the "critic" network) which is trained by SL to approximate the true value function for the current policy.

For an introduction to the different RL algorithms, the interested reader is referred to \cite{sutton2018book_rl}.

\subsection{Automatic differentiation and gradient-based optimization}

Deep learning is efficient because of the backpropagation technique that can efficiently calculate gradients with respect to all the hundreds or millions of parameters in a neural network. This technique more generally leads to the concept of automatic differentiation. There, the idea is to obtain the exact gradient with respect to any variable appearing in any kind of numerical calculation. As a numerical approach, this is distinct from symbolic differentiation applied in computer algebra programs. Modern frameworks used for neural networks offer various modes of automatic differentiation. This offers the chance to employ them for arbitrary gradient-based continuous optimization tasks, especially those involving many parameters, where efficiency is of concern. We will later show examples of how this can be used to discover new quantum experiments, quantum circuits, and in other contexts.

\subsection{How to get started}
Only a few years ago, implementing neural networks still required quite some effort. Nowadays, libraries like Tensorflow, PyTorch, JAX, and many more offer powerful tools to set up and train neural networks. We have created an online collection of resources to start with ML, that contains information on different popular frameworks, as well as helpful tutorials, lecture notes, and reviews for machine learning in general\footnote{We plan to update this collection continuously and are happy about contributions from the community.}: \url{https://github.com/ML4QTech/Collection}.

\section{Applications of Machine Learning for Quantum Technologies}
\label{sec:applications}

\subsection{Measurement data analysis and quantum state representation}

\begin{figure}
    \centering
    \includegraphics[width=\columnwidth]{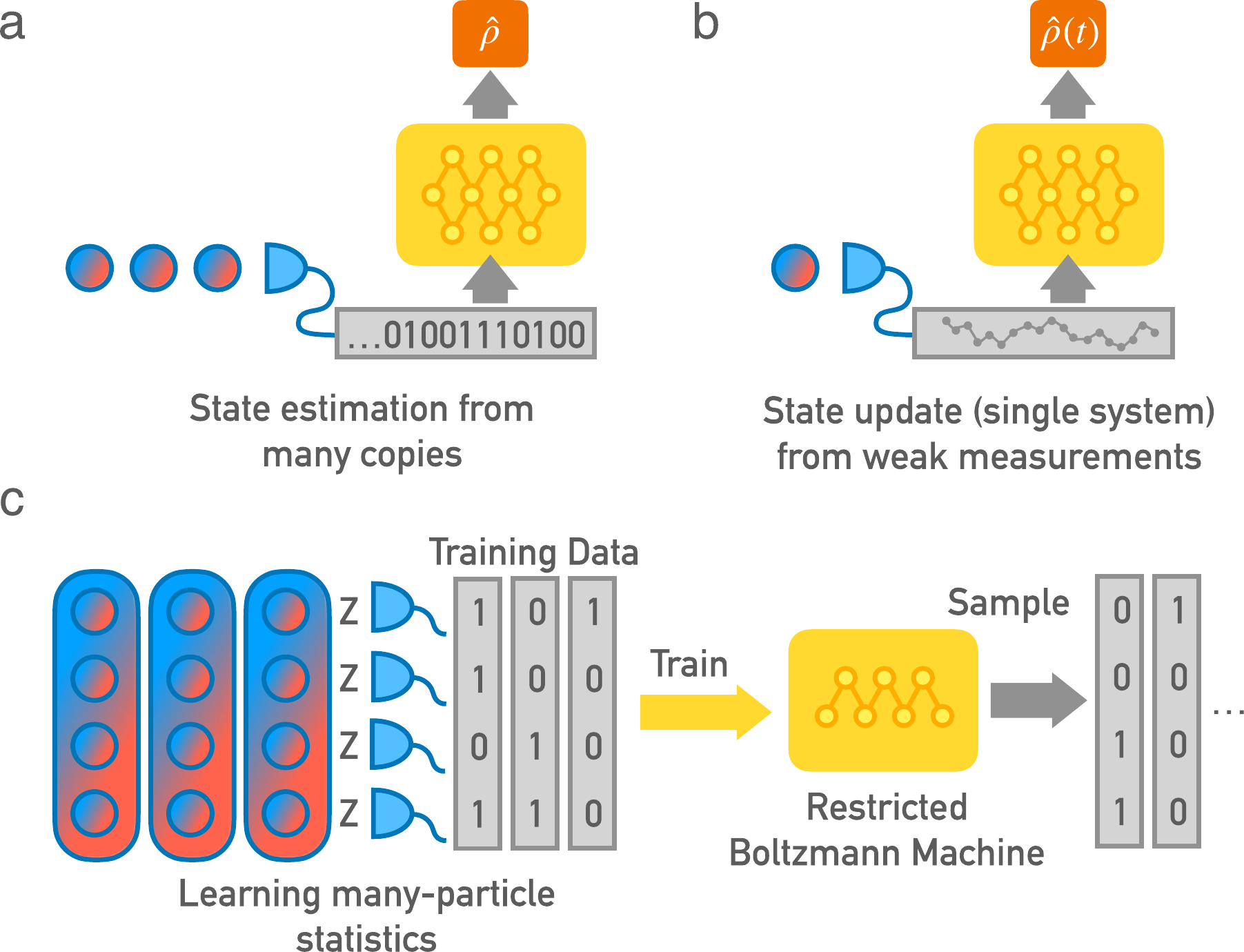}
    \caption{State Estimation via Neural Networks. (a) Measurements on many identical copies of a quantum state can be processed to produce an estimate of the quantum state.  (b) A continuous weak measurement on a single quantum system can be used to update the estimated state. Both in (a) and (b), a single network is trained to estimate arbitrary states correctly. (c) One can also train a network-based generative model to reproduce the statistics of a quantum state, i.e.~to sample from the probability distribution. Training requires many identical copies that can be measured, so the statistics can be learned. Here, one network represents only a single quantum state. It can be extended to handle measurements in arbitrary bases.}
    \label{fig:state_estimation}
\end{figure}

An important direct application of machine learning to quantum devices is the interpretation of measurement data. The application ranges from an improved understanding of the measurement apparatus itself to extracting some high-level properties of the quantum system to the full reconstruction of the measured quantum state. In many cases, this can be phrased as a supervised learning task. One example might be to extract an approximate description of a quantum state from a sequence of measurement results on identically prepared copies, see Fig.~\ref{fig:state_estimation}a. Provided the actual quantum state is known for each training example, the machine learning algorithm will learn to provide the best possible approximation to the quantum state. The same is true if the goal is to reconstruct certain properties of the state or the device instead of reconstructing the full quantum state. The choice of the cost function is essential, a simple example being the infidelity between the predicted and the actual state. Other choices will lead to slightly different optimal approximations predicted by the ML algorithm. Crucially, the algorithm can easily deal with distortions of the measurement data (such as extra technical noise), as it will learn the properties of these distortions and how to undo them. 

\textbf{Interpreting Measurements } -- An interesting early example used machine learning to improve the readout fidelity of a qubit in a superconducting quantum device. There, the noisy measurement trace, obtained from a microwave signal passing through a readout resonator interacting with the qubit, can be used to deduce the qubit's logical state. However, classifying the qubit state is challenging. The authors of \cite{magesan2015machine} use a basic machine learning technique called support-vector machine (SVM) to perform clustering of measurement traces in an unsupervised fashion, outperforming classical clustering algorithms. The idea of a nonlinear SVM is to map data points to a higher-dimensional space and to find the best hyperplane for separating two classes of data points in that space. In this specific work, each measurement trajectory, which consists of hundreds of individual data points, is interpreted as a point in a high-dimensional space. The SVM's goal was to separate curves that originate from a zero-state and one from a one-state. The readout fidelity was improved compared to non-ML clustering techniques. Furthermore, this analysis has shown that the main noise contribution comes from physical bit-flips (either heating or relaxation of the qubit). Without such events, the classification of the SVM becomes near-perfect. A similar approach has been demonstrated -- using neural networks -- to enhance the readout capability of trapped-ion qubits \cite{seif2018machine}. Here, the authors show that the readout fidelity improves significantly compared to a non-ML clustering method, especially when the effective amount of data per measurement increases. Similar techniques have also been applied to NV center quantum devices \cite{liu2020repetitive}.

In a pioneering experimental work, a neural network reconstructed the quantum dynamics of a quantum system directly from measurement data \cite{flurin_using_2020}. There, the authors considered again a superconducting qubit coupled to a readout resonator, whose noisy measurement trace is fed as input into the network (together with the initial preparation state and the final measurement basis). The network's task was to predict the statistics of arbitrary measurements at some given time during the evolution, i.e.~effectively predict the evolution of the density matrix given the measurement record, see Fig.~\ref{fig:state_estimation}b. The kind of network most suited to this task is a so-called recurrent network, i.e. a network able to process a time series (originally used for text or speech processing). The resulting fully trained network can map any measurement data to a quantum state.

Furthermore, machine learning techniques can be exploited to analyze the statistics of measurement outcomes in quantum experiments where the aim is to demonstrate the classical complexity of sampling from the quantum distribution. Boson sampling is the most well-known such scenario. In \cite{agresti_pattern_2019}, unsupervised machine learning techniques (various clustering approaches) were employed to identify and rule out possible malfunction scenarios that would lead to a noticeably different distribution.

One exciting work from theoretical computer science connects the question of quantum state tomography with the theory of computational learning in the supervised setting \cite{aaronson2007learnability}. Here, a learner uses the training dataset to produce a hypothesis about future measurements. Quantum state tomography, which requires a number of measurements that is exponential in the number of particles, can be seen as a learner that produces a hypothesis for every possible measurement on the quantum system. This might, however, not be necessary in most cases. The theory of \textit{probably approximately correct models} (PAC) provides a hypothesis for every measurement close to the dataset. Surprisingly, it was found that this question can be solved with a linear scaling of measurements. This computational learning strategy has been first demonstrated in quantum optics experiments with up to six photons. The experiment has confirmed the scaling behavior of PAC, even in the presence of realistic experimental noise \cite{rocchetto2019experimental}.

Another very interesting recent development shows that joint quantum measurements of several individual copies of a many-particle quantum state can lead to an exponential improvement over classical learning algorithms \cite{huang2022quantum}. The authors show experimentally on a platform of 40 superconducting qubits that tasks such as predicting properties of the physical system can be significantly improved when the results of such joint measurements are fed as an input to a classical RNN. The result is particularly remarkable as it shows a clear advantage already for current, noisy, and not error-corrected quantum computers.

Many other interesting examples exist that use neural networks to analyze simulated or measured data of quantum systems. For example, it has been demonstrated that the Wigner negativity of a multimode quantum state can be approximated well even in the low-data regime \cite{cimini2020neural}, with important consequences for quantum technologies. Another vivid field is the neural-network-based detection of quantum phase transitions and classification of quantum phases in condensed matter physics. We will not go into detail here. Instead, we point to some exciting early works in this field \cite{carrasquilla2017machine,van2017learning,wetzel2017unsupervised}, as well as some very modern applications that, for instance, use anomaly detection to find phase transitions in an unsupervised way \cite{kottmann2020unsupervised}. Anomaly detection has already been applied to detect phases directly from experimental measurements \cite{kaming2021unsupervised}. See a recent review on this topic in \cite{carrasquilla2020machine}.

\textbf{Approximation of Quantum States} -- The direct application of numerical techniques to quantum devices requires, in many situations, the storage, and processing of the system's full quantum state. As the quantum state grows exponentially with the number of particles, the memory requirements quickly become enormous, even for moderately large quantum systems. For example, storing the full quantum state of a 42-qubit system requires 35 TByte of memory. As demonstrated in the earliest quantum advantage experiments \cite{arute2019quantum}, this is directly related to the power of quantum computers and quantum simulators. However, it poses a significant problem for classical computational approaches that deal with large quantum systems and therefore for developing new large-scale quantum technologies.

To overcome this challenge, memory-efficient approximations of the quantum wave function are indispensable. Neural networks are one key candidate to approximate the quantum wave function. This approach has sometimes been called \textit{Neural Quantum State} (NQS).

A prominent approach tries to represent the quantum state in terms of a neural network \cite{carleo_solving_2017}. This implies that for each new quantum state, another network will be trained, based on the associated measurement data for that state. In principle, that is considerably easier than asking a single network to be responsible for arbitrary states, i.e.~the task considered above. As a consequence, much more complicated many-body states can be accessed. The whole approach can be seen as a neural-network-based version of quantum state tomography. 

Several different ways exist to use a single neural network to represent a single quantum state. One straightforward approach, first introduced in \cite{carleo_solving_2017} and then extended in subsequent works, employs a network that directly represents the wave function. Given a multi-particle configuration $x$ as input, the network has to produce the wave function amplitude for that configuration as output: $\Psi_{\theta}(x)$. 

Different structures can be used for the network, with a restricted Boltzmann machine (RBM) being a popular choice since it also allows direct sampling from the probability distribution of observations \cite{melko2019restricted}. In a traditional RBM, the aim is to learn to sample from some observed probability distribution, see Fig.~\ref{fig:state_estimation}(c). It consists of binary visible units and hidden units connected to each other, and the statistics of these units are sampled from a Boltzmann distribution with an energy $E$ that contains interaction terms bilinear in the hidden ($h$) and visible ($v$) unit values: $-E=\sum_j a_j v_j + \sum_k b_k h_k + \sum_{j,k} w_{jk} v_j h_k$. During training, the coupling constants are updated to obtain the desired probability distribution of $v$ (observed in samples provided during training). A simple physics example would be a 1D spin chain, whose configurations are identified as sample vectors $v$. More generally, other so-called generative deep learning methods (such as normalizing flows, variational auto-encoders, and generative adversarial networks) can be used to learn probability distributions, including those representing the statistics of observables in quantum states in a given basis.

Quantum state tomography using an RBM-style ansatz for the wave functions was introduced in \cite{torlai_neural-network_2018}. Since one wants to keep the wave function's phase $\varphi$ as well as the probability $p$, the ansatz is now of the type $\Psi(x)=\sqrt{p(x)} e^{i \varphi(x)}$, where both $p$ and $\varphi$ are represented as networks and $x$ corresponds to the visible units. A crucial idea in this approach is to match the probability distributions obtained from the experiment for observables in more than one basis (e.g.~$\sigma^z$ and $\sigma^x$ etc. for qubits). The evaluation of different bases can be carried out via unitary transformations acting on the wave function $\Psi$ that is expressed in a single reference basis. 

In \cite{torlai_integrating_2019}, this approach was applied to experimental data from snapshots of many-particle configurations taken in a Rydberg atom quantum simulator. The resulting network-based wave-function ansatz could then be used to reconstruct other expectation values and observables that were not directly accessed in the experiment.

The idea has been extended to mixed states \cite{schuld2019neural, nagy2019variational, hartmann2019neural, vicentini2019variational, yoshioka2019constructing}. In that way, NQS can be used to efficiently approximate open quantum systems which are notoriously difficult to capture.

{\bf Approximating Quantum Dynamics} -- Once a suitable quantum state representation is available, it can be exploited to evolve the state in time. This enables potentially efficient simulation of quantum many-body time evolution, which is important for predicting and benchmarking the dynamics of quantum simulators and quantum computing platforms. For the general case of dissipative quantum many-body dynamics, i.e. the time evolution of mixed states, this has been explored in \cite{hartmann2019neural}. 

Rather than explicitly storing the entire quantum state, another technique shows how one can directly compute a quantum state's complex properties just from the state's construction rules, i.e.~the quantum experimental circuit. In \cite{adler_quantum_2021}, the authors show how a recurrent neural network \cite{hochreiter1997long} can approximate the properties that emerge from quantum experiments without ever storing the intermediate quantum state directly. These systems could then directly be applied for complex quantum design tasks, a topic we cover in chapter \ref{Chapter_DesignQuantumExperiment}. Another approach \cite{mohseni2022deep} also foregoes the representation of quantum states and instead trains a recurrent network to predict the evolution of observables under random external driving of a quantum many-body system (either based on simulated or possibly even experimental data). The trained network can then predict the evolution under arbitrary driving patterns (e.g.~quenches). In a similar spirit, neural ordinary differential equations (neural ODEs) can be used to approximate the dynamics of quantum systems directly, again without storing the explicit information about the quantum wave function \cite{choi2022learning}. Interestingly, the approximation is of high enough quality that it is possible to rediscover some fundamental properties of quantum physics, such as the Heisenberg uncertainty relation.

\textbf{Future Challenges and Opportunities} --

Improved data efficiency (both for the training but also when applying ML to interpret the data) will be an important challenge for the future, especially when the devices scale to more complex quantum systems.

It will be interesting to co-discover measurement strategy together with the data interpretation strategy. This might be particularly interesting if the AI algorithm is allowed to employ quantum measurements on numerous copies of the same state, as pioneered in \cite{huang2022quantum}.

When a neural network can find a suitable approximation for the computation of complex quantum systems, such as an NQS, it has learned a theoretical technique that might be useful for humans too. It will be interesting to learn how to extract the per-se inaccessible knowledge from the weights and biases of the neural network. One method is so-called symbolic regression.

The extensions of NQS to complex quantum systems, such as higher dimensions and spins beyond qubits, will allow for more interesting applications.

\subsection{Parameter estimation: learning the properties of quantum systems}
In this section, we will discuss machine-learning-based approaches for estimating experimental parameters. These could either be system parameters for the calibration of quantum devices or the estimation of external parameters in the form of quantum metrology.

\begin{figure}
    \centering
    \includegraphics[width=\columnwidth]{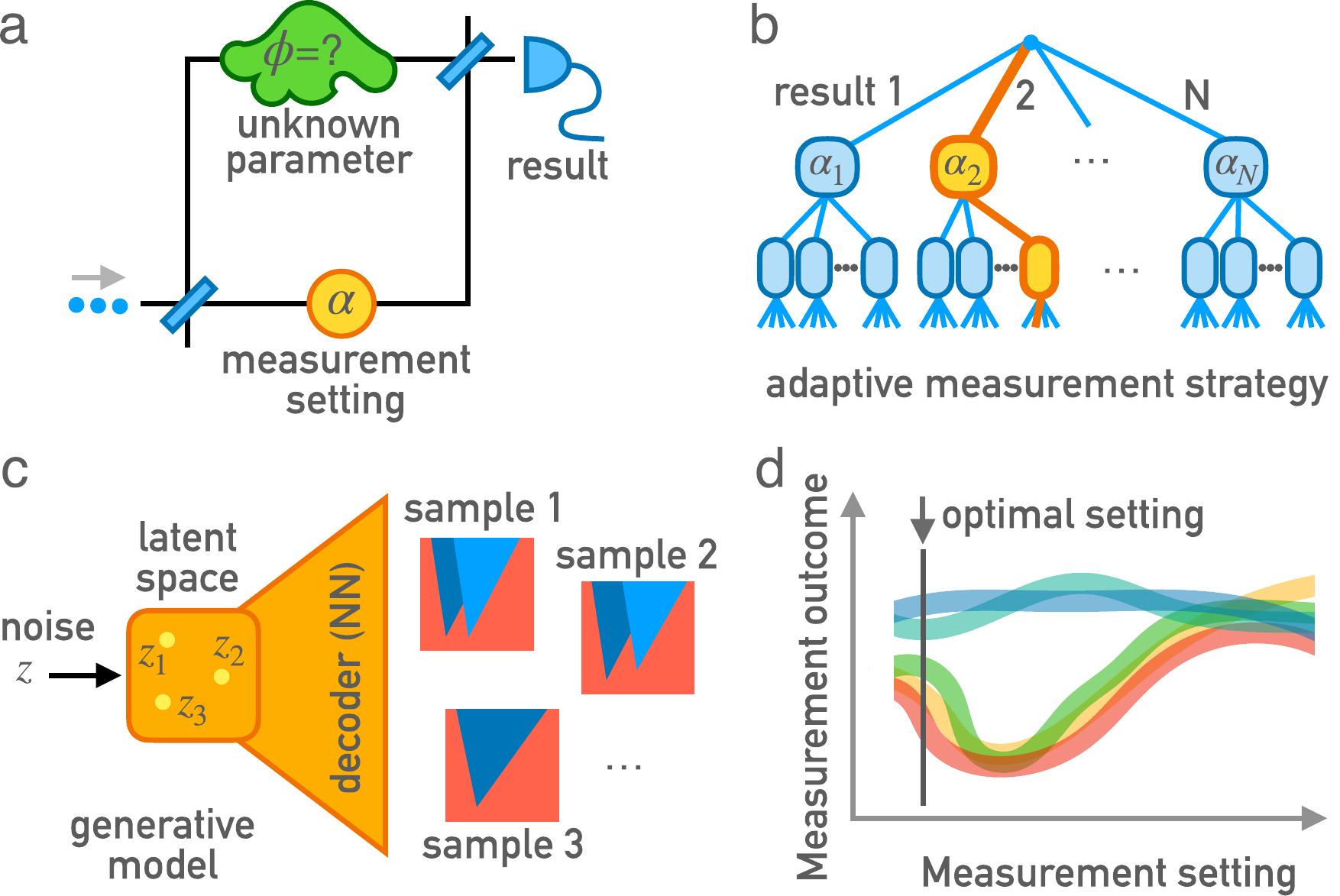}
    \caption{
    Machine Learning for Parameter Estimation in Quantum Devices. (a) A typical scenario, with the measurement result statistics depending both on some tuneable measurement setting and the unknown parameter(s), here represented as phase shifts in a Mach-Zehnder setup. (b)
    An adaptive measurement strategy can be illustrated as a tree, with branches on each level corresponding to different measurement outcomes. Depending on those outcomes, a certain next measurement setting (indicated as "$\alpha_j$") needs to be selected. Finding the best strategy is a challenging task, as it corresponds to searching the space of all such trees. (c) Neural generative models can be used to randomly sample possible future measurement outcomes (here: 2D current-voltage maps as in \cite{lennon2019efficiently}) that are compatible with previous measurement outcomes. This is helpful for selecting the optimal next measurement location. Different random locations in latent space result in different samples. (d)  Measurement outcome vs.~measurement setting for 5 possible underlying parameter values (different curves; measurement uncertainty indicated via thickness). We aim to maximize the information gain, i.e.~choose the setting which best pinpoints the parameter (which is not equivalent to maximizing the uncertainty of the outcome). 
    }
    \label{fig:metrology}
\end{figure}

\textbf{Quantum metrology} -- Machine learning can be helpful for various challenges in quantum metrology \cite{polino2020photonic}. Quantum metrology deals with the resource-efficient measurement of external parameters acting on a quantum system, like magnetic fields or optical properties of a material sample. Broadly, the field can be separated into two different branches. First, non-adaptive approaches exploit how complex quantum entanglement can be exploited to reduce the required resources (for instance, the number of photons that interact with a sample) without changing the experimental setup or input states throughout the measurement sequence. The second class of approaches exploits feedback, meaning they employ adaptive strategies that change either the input or the measurement setting, depending on the previous measurement outcome. Naturally, adaptive strategies require more advanced experimental implementations, including fast switching or long-term stability of setups. For that reason, non-adaptive quantum metrology is so far much more explored in laboratories, while adaptive approaches are still at the stage of proof-of-principle experiments. AI has contributed to both approaches. 

An example of the application of neural networks to the estimation of unknown parameters in a photonic experiment is provided in \cite{cimini2019calibration}. There the goal is to calibrate a device via neural-network training that can later be used to estimate an unknown phase shift. The authors first accumulate a large set of calibration measurements using a controlled phase plate. In addition, the data is augmented to account for the statistical noise contribution. The neural network receives the measurement data (the number of detected photons) and has to estimate the corresponding phase. As soon as the neural network is able to model the connection between measurement data and calibration phases, it can be used to estimate unknown phases as well. This task shows that device calibration and estimation of external parameters for quantum sensors are closely related.

In addition, also still for non-adaptive approaches, various projects have explored the discovery of new experimental setups for the resource-efficient measurement of parameters of an external system \cite{knott2016search, nichols2019designing}. We will talk about this approach in section \ref{Chapter_DesignQuantumExperiment}.

We now turn to focus on feedback-based quantum metrology schemes and how they could be improved with ML algorithms. A pioneering non-ML approach for such a task is called BWB (Berry-Wiseman-Breslin) strategy, after the authors of
\cite{berry2000optimal,berry2001optimal}. In the original setting of that strategy, the authors consider a Mach-Zehnder interferometer with two detectors at the two outputs, see Fig.~\ref{fig:metrology}a. One of the arms contains an unknown phase shift, and the other arm contains a controllable phase that is modified depending on the previous measurement results. One can think of this decision-making problem as a decision tree (see Fig.~\ref{fig:metrology}b), where the actions must be chosen based on the measurement outcomes to maximize the information gain. In general, the expected information gain is defined as the reduction of entropy of the parameter distribution, averaged over possible measurement outcomes. The authors derived the feedback algorithm by applying Bayes theorem to the distribution of the unknown phase, which is then used to choose subsequent measurement settings with a large information gain. This strategy has been found not only to be highly resource-efficient, but also practically implementable in the laboratory \cite{higgins2007entanglement}.

Interestingly, greedy adaptive strategies (i.e.~to choose in every step the reference phase that yields the largest immediate information gain) do not necessarily lead to the largest information gain in the long run. To overcome this effect, one of the seminal early contributions to ML-based quantum metrology used particle swarm optimization of the feedback strategy\cite{hentschel2010adaptive_phase_estimation}. In particle swarm optimization, a collection of different feedback strategies (each called ``particle'') iteratively moves in the space of all possible strategies. In each iteration, the particle moves towards a combination of the best local optimum and the currently best known global optimum known by the whole swarm. The experimental setting of \cite{hentschel2010adaptive_phase_estimation,hentschel2011adaptive_phase_estimation} is the same as for the BWB strategy -- a Mach Zehnder interferometer with an unknown phase in one arm and an adaptive phase in the other arm. Indeed, the swarm optimization algorithm finds (slighly) better strategies than the greedy Bayesian BWB approach. Interestingly, neither BWB nor swarm optimization can find the optimal strategy, which was identified for small photon numbers via an extensive computation of all possible strategies. Other early ML algorithms in this domain have applied evolutionary approaches to approximate the ideal feedback strategy\cite{lovett2013adaptive_phase_estimation,palittapongarnpim2017genetic_algo_for_quantum_control_and_metrology}.

Discovering a strategy is a problem that can directly be formulated as a reinforcement learning task. An early application of RL to quantum parameter estimation was provided in \cite{xu2019generalizable}, with frequency estimation of a qubit as a test case. In that work, the idea was to optimize the quantum Fisher information for the parameter of interest. This can be done by finding a sequence of suitable control pulses applied during the noisy evolution of the quantum probe. No feedback is involved in this simple setting since the measurement itself is not part of the evolution controlled by RL. 

However, the quantum Fisher information is only useful in cases where one is already fairly certain of the true parameter value. RL can be employed to study more complex situations, where updates are performed using the Bayes rule, starting from an arbitrary prior parameter distribution, and where the strategy is not greedy (i.e.~more than a single step of the sequence is optimized). In \cite{schuff2020improving}, the authors provided information about the current Bayes distribution of the unknown parameter (as extracted from previous measurement results) and the previous measurement choices as input to an RL agent implemented by a neural network. It then has to suggest the next measurement. After the whole sequence of measurements, the agent is rewarded according to the total reduction in parameter variance. It was shown that this approach performs very competitively for an important test case, namely parameter estimation for a qubit of unknown frequency in the presence of dephasing.

\textbf{Device calibration} --
Future large-scale quantum devices will consist of a large number of components with adjustable parameters that need to be characterized and tuned automatically. A complete characterization of the device via quantum process tomography quickly becomes impractical. To find the actual parameters or the ideal operating point of the quantum device, it is therefore necessary to extract the relevant data with a very limited amount of information. This task can be formulated as a machine learning task, specifically applying ``active learning'' or ``Bayesian optimal experimental design'' \cite{ryan2016review} where the algorithm chooses the most informative measurement autonomously (see also \cite{gebhart2022learning} for a review of such methods in the context of quantum devices). Naturally, this is closely related to the estimation of external parameters in quantum metrology as discussed above.

We illustrate these techniques via a pioneering experimental application to quantum devices. This experiment \cite{lennon2019efficiently} considered the calibration and measurement of a semiconductor quantum dot. Such a device can be tuned via applied gate voltages, and its resulting properties can be measured via a transport current. Here, the goal was to explore the properties of the quantum dot, as defined by its current-voltage map $I(V_1,V_2)$, where the voltages include a bias voltage driving the current and a gate voltage deforming the dot's potential. In a naive approach, even if only two voltages were scanned with 100 discretization steps each, one would need to perform 10,000 measurements to get a suitable resolved device characteristic. 

To reduce the required number of current measurements, the authors tried to estimate which measurement (in the 2D voltage space) would yield the "maximum amount of additional information." In practical terms, this would be the measurement that is expected to place the tightest constraints on the current-voltage maps that are still compatible with all the observed values of the current (observed in this and prior measurements). It is obvious how this setting translates to other quantum platforms, e.g.~measurements of microwave transmission through superconducting circuits controlled via gate voltages and magnetic fields or the optical response of tuneable atomic systems.

The first step towards this goal is to efficiently represent all current-voltage maps that might be observed, given the general physics of such a device, the assumed prior distribution of device parameters, and all previous measurement results. In general, this is the domain of "generative models," which can sample from a probability distribution that is learned. In the case of \cite{lennon2019efficiently}, the authors used such a generative model, in their case a so-called "constrained variational autoencoder" (cVAE), to randomly create realistic current-voltage map that follow the probability distribution of the actual physical system, see Fig.~\ref{fig:metrology}c. Additional input into the generative model provides a constraint, in the form of a few initially existing measurement results, and guides the reconstruction to sample only maps compatible with those constraints. In each step, 100 different voltage maps are sampled. Those maps are used to find the next measurement point in voltage-space that would lead to the maximum information gain, see Fig.~\ref{fig:metrology}d. With this technique, the total number of necessary measurements for the characterization of the device is reduced by a factor of 4. This clearly shows that the overhead of the deep-learning algorithm is more than compensated by its efficiency improvement compared to the naive approach. A benefit of this technique is that generating new samples with the cVAE is very efficient. Thus it can be scaled to much larger devices, where even more significant efficiency gains are expected.

Another comparatively straightforward way to use machine learning in device characterization consists in training a network-based classifier to recognize "interesting" measurement results. This then allows to tuning parameters until those results are obtained. Such an approach has been demonstrated in \cite{durrer_automated_2020} for navigating charge-stability diagrams of multi-quantum-dot devices. In that setting, the algorithm's goal was to automatically tune the charge occupation of the double quantum dot. The task is reformulated as a classification task, where the algorithm recognizes individual charge transitions when presented with a charge-stability diagram. Since such a diagram constitutes an image, CNNs are a suitable choice for the task.

\textbf{Quantum Hamiltonian Learning} -- Imagine the following parameter-estimation problem: One wants to estimate the parameters $x_0$ that affect the evolution of a quantum state under a {\em quantum many-body} Hamiltonian $H(x_0)$ \cite{wiebe2014hamiltonian}. Unfortunately, even the task of computing the dynamics scales exponentially with the system size (number of qubits) when tackled using a classical machine. The idea of Quantum Hamiltonian Learning (QHL) is to enlist the help of a quantum simulator to overcome this problem. The parameters $x_0$ can then be estimated with standard Bayesian methods. Thereby, the quantum simulator is used like a subroutine inside a classical ML approach. The first experimental implementation of this idea was demonstrated in 2017 \cite{wang2017experimental}. In that work, the authors wanted to estimate the parameters of an electron spin in a nitrogen-vacancy center, and they used a quantum simulator on an integrated photonics platform to perform the QHL. Interestingly, not only did the approach lead to a high-quality estimation of the dynamic system parameters, but it also indicated when the initial Hamiltonian model had deficits. In these cases, the learning method informed the user that there are other dynamics in play that have not been considered, which inspires an improvement of the underlying Hamiltonian model.

While the QHL method indicates that when the model Hamiltonian is not ideal, it cannot adapt it. To overcome this hurdle and to learn the entire Hamiltonian structure (not only its parameters), the authors of \cite{gentile2021learning} have introduced the idea of a \textit{Quantum Model Learning Agent} (QMLA). This agent not only finds the parameters of a predefined Hamiltonian, but discovers the whole Hamiltonian that describes the dynamics of a system. The approach iteratively refines the initial Hamiltonian and uses QHL as a subroutine for finding suitable parameter settings. This approach has also been demonstrated in a hybrid quantum system involving a nitrogen-vacancy center. The underlying learning mechanism is very general, and it thus could become a powerful tool for learning the dynamics of unknown quantum systems.

\textbf{Future Challenges and Opportunities} --
For adaptive approaches, one needs to consider the trade-off between speed and the sophistication of the approach. In these tasks, the time between measurement and feedback is often very short, so the decision must be taken quickly.

An interesting future approach for advanced quantum metrology approaches is to simultaneously co-design the experimental setup and the feedback strategy, rather than solving these tasks individually.

\subsection{Discovering strategies for hardware-level quantum control}
\label{sec:hardware-level-control}

Challenges like quantum computing and quantum simulation are leading to rapidly increasing demands on the efficient and high-fidelity control of quantum systems. Tasks range from the preparation of complex quantum states and the synthesis of unitary gates via suitable control-field pulses all the way up to goals like feedback-based quantum state stabilization and continuously performed error correction. In trying to solve these tasks, the specific capabilities and restrictions of any hardware platform, from superconducting circuits to cavity quantum electrodynamics, need to be considered.

\begin{figure}
    \centering
    \includegraphics[width=\columnwidth]{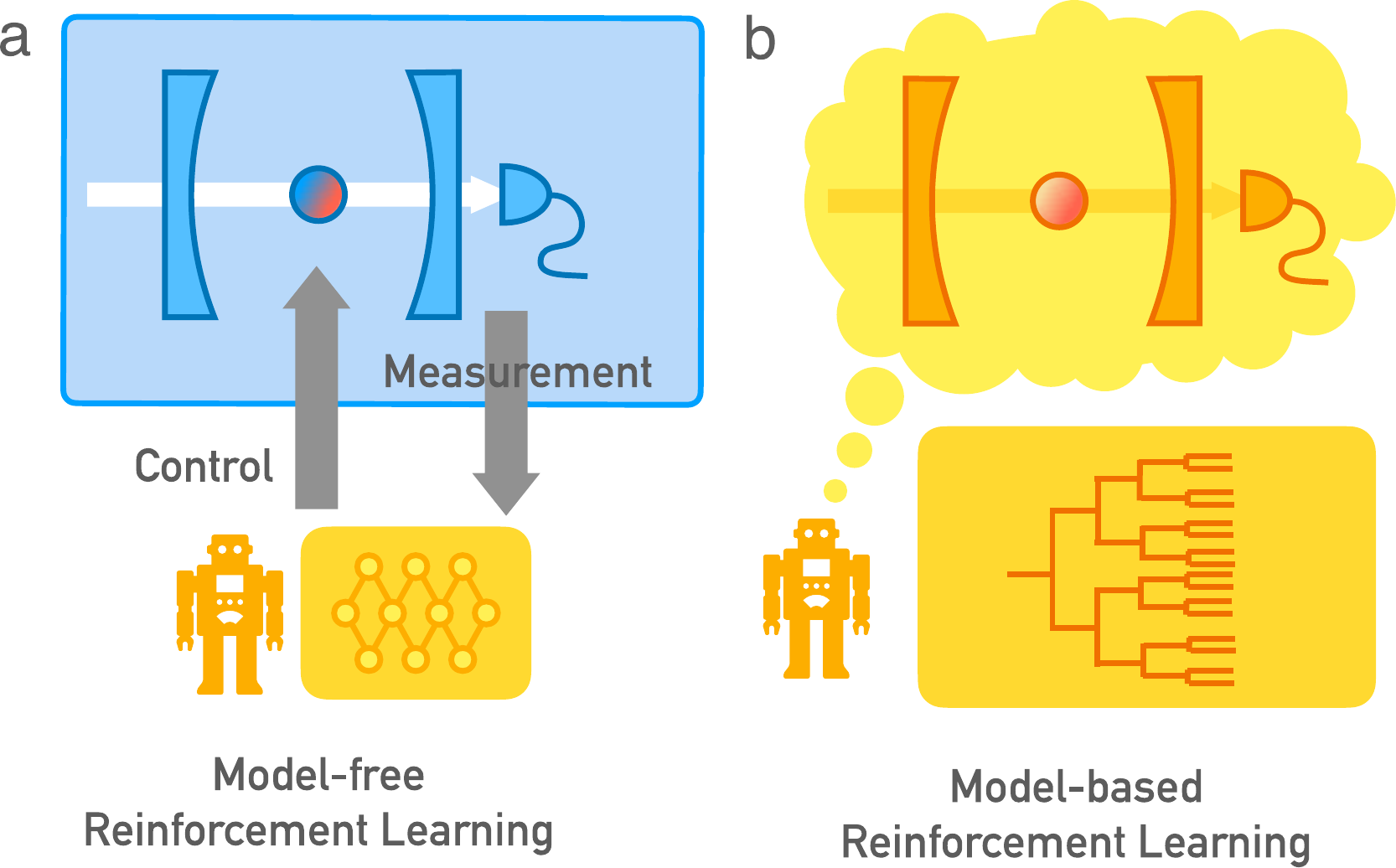}
    \caption{(a) The eventual goal of model-free reinforcement learning is the direct application to experiments, which then can be treated as a black box. Many actual implementations, however, use model-free RL techniques applied to model-based simulations. (b) Model-based reinforcement learning directly exploits the availability of a model, e.g.~taking gradients through differentiable dynamics.}
    \label{fig:modelfree_modelbased}
\end{figure}

In this section, we will highlight specifically how reinforcement learning has come to help with many of these challenges. In the form of model-free RL, it promises to discover optimal strategies directly on an experiment, which can be treated as a black box, see Fig.~\ref{fig:modelfree_modelbased}a. All its unknowns and non-idealities will then be revealed only via its response to the externally imposed control drives. But even when used in a model-based way, using simulations, RL can be more flexible than simpler approaches. In particular, it offers ways to discover feedback strategies, i.e.~strategies conditioned on measurement outcomes. These were not previously accessible to the usual numerical optimal control techniques.

The present section is firmly concerned with hardware-level control that is continuous in the time domain, discovering pulse shapes or feedback strategies based on time-continuous noisy measurement traces as they would emerge from weak measurements of quantum devices. There are some connections to the next section, but there we will be concerned with the discovery of protocols, control strategies, and whole experimental setups that are described on a higher level, composed of discrete building blocks like gates or experimental elements.

\textbf{Quantum control tasks without feedback (open-loop control)} -- 

Prior to the application of machine learning techniques in this field, the focus was essentially on tasks without feedback, which was solved by direct optimization techniques, adapting the shape of control pulses applied to the quantum system to maximize some quantity (like the state fidelity). These direct optimization techniques include gradient-based approaches, with GRAPE \cite{khaneja2005optimal} and the Krotov method \cite{krotov1995global,somloi1993controlled} the most prominent examples, as well as approaches that do not rely on access to gradients, such as CRAB \cite{doria2011optimal}. At the time of writing, these techniques still form the default toolbox for the case of open-loop control, even while the first applications of machine learning (described below) are taking hold. Evolutionary algorithms define another class of (stochastic) approaches that have been used successfully to find optimal control sequences \cite{judson1992teaching}.

State preparation is the most common quantum control problem, and yet it can already be challenging, especially for multi-qubit settings. In probably the earliest application of RL to quantum physics, pure-state preparation with discrete control pulses was shown using a version of Q-learning for a spin-1/2 system and a three-level system \cite{chen2013rl_for_quantum_control}. A few years later, RL-based state preparation was demonstrated for a many-qubit system \cite{bukov_reinforcement_2018}, also using Q-learning and discrete bang-bang type actions, with particular emphasis on analyzing the complexity of the control problem showing up in the form of a glassy control landscape. Both of these works used some version of table-based Q-learning, which works well for a restricted number of states and actions. The first work to employ deep (i.e. neural-network-based) RL methods to open-loop control of quantum systems was \cite{august2018taking}, with both discrete and continuous controls and a recurrent network as an agent, as applied to dynamical decoupling and again state preparation, followed shortly afterwards by \cite{zhang2018automatic}.

The RL approach can be used successfully to find suitable pulse sequences for unitary gates and optimize for the gate fidelity, as shown first in \cite{niu_universal_2019}, and analyzed later also in \cite{an2019deep}. 

Recently, deep RL has been applied for the first time to learn control strategies for a real quantum computing experiment \cite{baum_experimental_2021}. The authors trained on a cloud-based quantum computing platform, collecting data for the current control policy, extracting rewards, and updating the policy. The goal was unitary gate synthesis, and the lack of real-time access to the device was not a concern since the task required only open-loop control. This first demonstration of RL-based quantum control on a real quantum experiment helped to illustrate the possibilities and challenges in this new approach.

Even though open-loop control pulse design means that the actual strategy in the experiment is not conditioned on any measurement outcomes, RL training for such tasks (when done on a computer simulation) may still benefit from the agent receiving input information like the current quantum state. Experience shows that this makes it easier to find a good strategy. Otherwise, only very sparse nominal information like the current time step and possibly the most recent selected action would be fed into the agent. In any case, however, once RL has found a control sequence, it could in principle be stored (e.g. as a waveform or pulse sequence) and sent to an experiment whenever needed. In other words, there is no need for the agent to be running during the actual experiment, which strongly relaxes requirements for the hardware: no real-time control is necessary.

{\bf Quantum feedback control (closed-loop control)} -- The successful control of quantum systems subject to noise, decay and decoherence requires either reservoir engineering (autonomous feedback) or active feedback control. The space of active feedback strategies is exponentially larger than that of open-loop control strategies (i.e.~without feedback), owing to the number of potential measurement outcome sequences growing exponentially with time (each such sequence may require a different response). It is here that it is almost inevitable to use the power of RL, particularly deep RL, with its ability to process high-dimensional observations.

The first work to apply deep RL to {\em feedback-based} control of quantum systems was \cite{fosel_reinforcement_2018}. It employed discrete gates for quantum error correction, and we will discuss it in sections \ref{Chapter_DesignQuantumExperiment} and \ref{sec:Quantum_Error_Correction}. State preparation and stabilization in the presence of noise or an uncertain initial state are other natural candidates for RL feedback strategies. Examples include quantum state engineering via feedback \cite{mackeprang2020reinforcement}, as well as control of a quantum particle in an unstable potential \cite{wang2020deep} and a double-well potential \cite{borah2021measurement}. In some quantum systems, control may be very limited (e.g. only linear manipulations), but measurements can introduce nonlinearity and their exploitation through RL-based feedback strategies can enable powerful control, as shown in \cite{porotti2021deep}. One challenge for model-free RL as applied to experiments is to make sure rewards can be extracted directly and reliably from experimental measurements, and to use a training procedure that really treats the quantum device as a black box (not relying, e.g.~on simulations). These aspects were emphasized in \cite{sivak2022model}, where state preparation in a cavity coupled to a qubit was analyzed. 

{\bf Model-free vs model-based RL} -- Applying model-free RL techniques, as described above, has a great advantage: the experimental quantum device can be treated as a black box, and its inner parameters and distortions of the control and measurement signals need not be known a priori. However, this also means that part of the training effort is spent on effectively learning an implicit model of the quantum device since that is the basis for a good control strategy. 

In many situations relevant to modern quantum technologies, though, a good model is known since the Hamiltonian and Lindblad dissipation terms have been carefully calibrated. This allows to consider model-based techniques explicitly, see Fig.~\ref{fig:modelfree_modelbased}b. In principle, these can simply consist in applying model-free approaches to an RL-environment that is represented by a simulation of the model. However, this is only useful if running the experiment often would be expensive or time-consuming. A more direct approach takes gradients directly through the model dynamics. In the absence of feedback, this is what well-known approaches like GRAPE offer. In an interesting recent development, automatic differentiation (the cornerstone of deep-learning frameworks) has been used to easily get access to the gradients needed for model-based control optimization. This was first presented in \cite{leung2017speedup} and then applied to various quantum control tasks, especially for qubit systems \cite{abdelhafez2019gradient,abdelhafez2020universal}, also employing neural networks to generate the control pulses \cite{schafer2020differentiable,coopmans2021protocol}.

Until recently, it was unclear, however, how to extend these ideas naturally to situations with feedback. The reason is that the stochastic choice of measurement outcomes is not directly compatible with taking gradients through smooth dynamics, unless special care is taken. A first example, applied to feedback based on weak linear measurements, was provided in  \cite{schafer2021control}, using automatic differentiation. The continuous measurement outcomes in such situations can be written as a simple function of a given Gaussian noise process. Very recently, a fully general approach, termed `feedback-GRAPE,' was presented and analyzed in \cite{porotti2022gradient}. There, it was pointed out how the effect of discrete stochastic measurements can be properly considered in such a gradient-based setting. This enables model-based optimization for feedback involving arbitrary strong discrete measurements, where the response to those outcomes can be represented via neural networks or trainable lookup tables.

\textbf{Future Challenges and Opportunities} -- As of the time of writing, experimental applications of reinforcement-learning-based quantum control are still in their infancy. Even for the easier, open-loop control case, one has to set up a full pipeline where control sequences are delivered to the setup and a suitable reward is extracted from experimental measurement data, before being processed (e.g.~externally, in a PC, implementing the RL algorithm). In a quantum system, where measurements collapse the state, this often means the reward can only be obtained at the end of an experimental run, making it harder to guide training. By contrast, having an immediate reward after each time step would improve training success because it assigns credit to the actions that immediately preceded a high reward. 

The challenge becomes even larger when feedback control is called for. Then, we require an agent (a neural network) that can process in real time the incoming measurement signals and decide on the subsequent actions. Depending on the hardware platform, this imposes severe constraints: for superconducting qubits, the time afforded to one such evaluation may be on the order of only a few hundred nanoseconds. This kind of challenge is specific to real-time feedback and does not exist for any of the other machine learning applications to quantum devices.

Almost all of the applications of model-free RL techniques mentioned above are numerical and rely on some simulation of the quantum device (if only because realizing these approaches in an experiment is still technically very demanding). In many publications, access to the simulation is furthermore utilized to make learning easier, e.g. by feeding the current quantum state obtained from the simulation as an ``observation'' into the agent. Since that would not be available in a real experiment, one possible solution is the use of a stochastic master equation to deduce the quantum evolution of the state based on noisy measurement traces during the experimental run in real time (see e.g. \cite{porotti2021deep} for comments on this). Another option is a ``two-stage learning'' procedure, where the successful RL agent, which still uses the state as input, is used for supervised training of another network that only uses measurement results as input. This new network can then be applied to an experiment \cite{fosel_reinforcement_2018}. However, both approaches require some calibration of experimental parameters since the original training still relies on simulations. 

\subsection{Discovering quantum experiments, protocols, and circuits}\label{Chapter_DesignQuantumExperiment}
\begin{figure}
    \centering
    \includegraphics[width=\columnwidth]{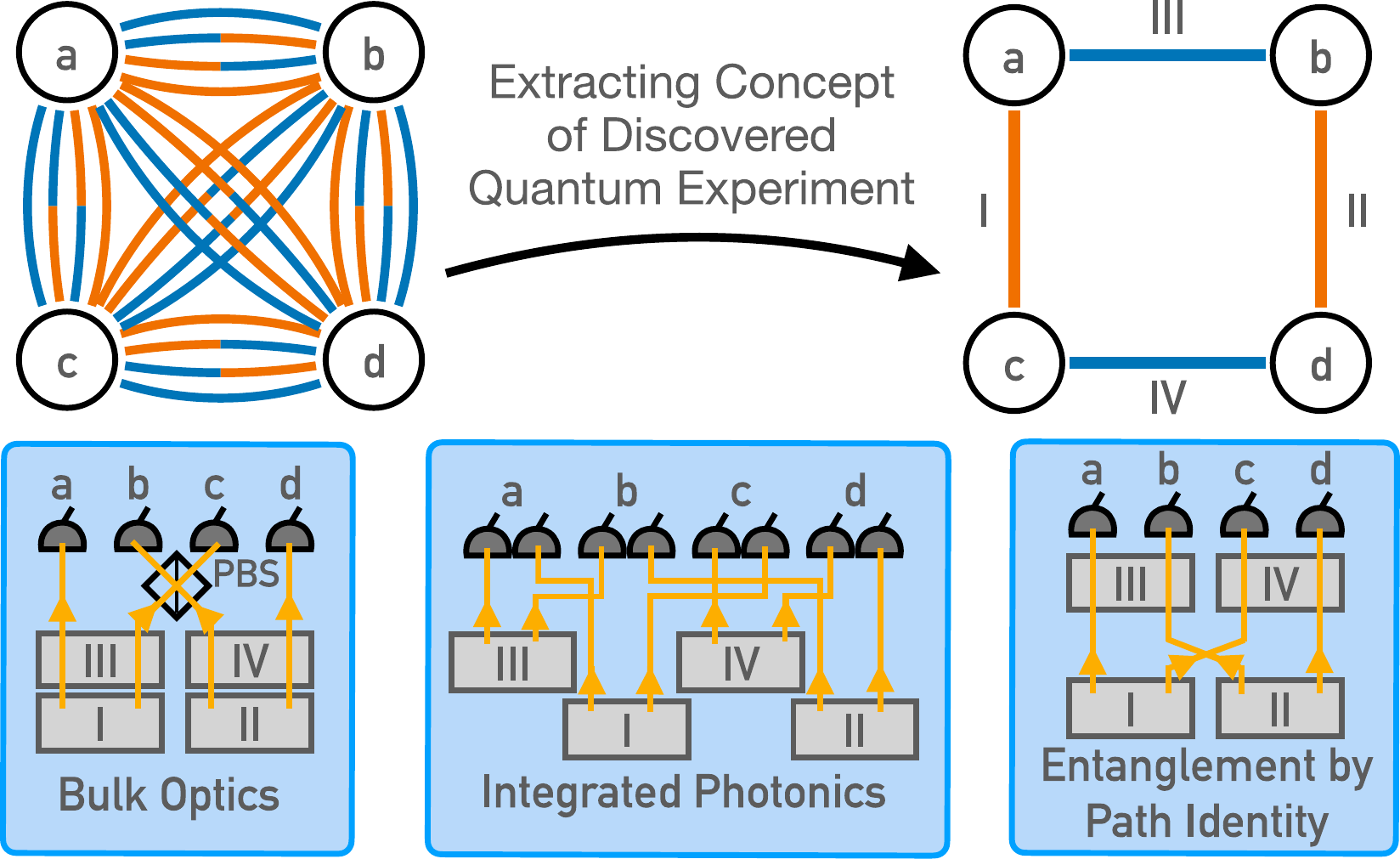}
    \caption{Discovery of Quantum Experiments. Quantum optics experiments can be represented by colored graphs. Using the most general, complete graph as a starting representation, the AI's goal is to extract the conceptual core of the solution, which can then be understood by human scientists. The solution can then be translated to numerous different experimental configurations\cite{krenn2021conceptual}. }
    \label{fig:discover_quantum_experiments}
\end{figure}
Inventing a blueprint for a quantum experiment, circuits or protocols requires ensuring that complex quantum phenomena play together to produce a quantum state or a quantum transformation from a limited set of basic building blocks. In this section we will discuss design questions that involve discrete building, such as optical elements, superconducting circuit elements or quantum gates. While these systems also contain additional continuous parameters, the overarching discrete nature gives in some cases the possibility -- as we will see -- to \textit{understand} and learn from the solutions.

\textbf{Discovery of Quantum Experiments} -- An important and natural playground for discovery based on discrete building blocks is the invention of new experimental setups.

In the field of quantum optics, those building blocks contain lasers, nonlinear crystals, beam splitters, holograms, photon detectors, and alike. Put together in a specific way, those systems can lead to the generation of complex quantum entanglement or perform quantum teleportation -- but exchange one single element, and the setup produces something entirely different and most likely not useful. Conventionally, experienced and creative scientists use their intuition and insights to design new quantum experiments. They translate an abstract task into a concrete layout that can be built in an experimental laboratory. However, human researchers are struggling to find suitable experimental setups for more complex quantum states and transformations. 

This challenge was met recently with the introduction of automated discovery of quantum experiments from scratch\cite{krenn_automated_2016,krenn_computer-inspired_2020}. The general idea is that an algorithm combines building blocks from a toolbox to produce a suitable experimental setup while optimizing for certain desired characteristics. In the setting of \cite{krenn_automated_2016,krenn_computer-inspired_2020}, the algorithm (here called \textit{Melvin}), puts optical elements from a toolbox onto a virtual optical table. The toolbox consists of the experimental components available in the laboratory (such as lasers and crystals). The algorithm initially starts to put elements from the toolbox in random order on the table. If the candidate setup satisfies all sanity checks, a simulator computes the full experimental output. If the setup produces the desired quantum state, it is automatically simplified and reported to the user. In addition, the setup is then stored as a new part of the toolbox. In that way, over time, the algorithm learns useful macro components which it can use in subsequent iterations. Thereby, it can already access useful operations that significantly speed up the discovery process.

This algorithm has produced experimental blueprints that enabled the observation of numerous new quantum phenomena in laboratories \cite{krenn_computer-inspired_2020}. Furthermore, new concepts and ideas have been discovered, understood, and generalized from some of the surprising solutions of the algorithm, such as an entirely new way to multi-photon interactions \cite{gao2020computer}.

Sequentially building an experimental setup can be formulated as a reinforcement learning problem. This possibility has been explored in \cite{melnikov2018active}. The approach led to the rediscovery of several experimental setups and to the automated simplification of experimental setups (which before was done using hand-crafted algorithms). 

An alternative approach, called \textit{Theseus}, which is orders of magnitudes faster than the previous techniques, is based on a new abstract representation of quantum experiments \cite{krenn2021conceptual}. Here, quantum experimental setups are translated into a graph-based representations, see Fig.\ref{fig:discover_quantum_experiments}. Any quantum optical experiment that can be built in the laboratory can be represented with a colored weighted graph, which translates an in principle infinite search space into a finite space with continuous (thus differentiable) parameters. In that way, the question of finding a certain quantum setup can be directly translated into discovering a graph with certain properties. As the parameters are continuous, highly efficient gradient-based optimization algorithms can be used to find the solutions. In addition, the graph's topology is eventually  simplified, such that the human researcher is not only presented with a solution, but can immediately understand \textit{why} and how the solution works. The algorithm has been used to answer numerous open questions and has led to new concepts. It showcases that -- when a simulator is available (i.e. a model) and no feedback needs to be taken into account, gradient-descent optimization at a large abstract representation often outperforms approaches such as genetic algorithms networks or reinforcement learning.

A different approach, denoted \textit{Tachikoma}, aims at the discovery of new experimental setups for quantum metrology\cite{knott2016search,nichols2019designing} -- using an evolutionary learning approach. The goal of \textit{Tachikoma} is to find setups for quantum states that can measure phase shifts efficiently and with high precision. It uses a toolbox of optical elements from which it builds up a pool of candidate solutions. The next generation is produced from the best-performing parent setups. Those are merged and mutated to create the new pool of candidates. In that way, the population improves its performance over time and leads to numerous counter-intuitive and exotic solutions. One of the computationally expensive operations and bottlenecks is the computation of the fidelity of the quantum state. For that, the authors have extended the approach by using a neural network that can classify the quantum states from the setup. This combination of an evolutionary algorithm with a neural network has discovered experimental blueprints with yet unachieved quantum metrology advantage. While much effort has been put into constraining the system to realistic solutions, it remains to be seen whether these experiments can be built in the laboratory and achieve the expected quality.

An entirely different approach uses logical artificial intelligence for designing quantum optical experiments \cite{cervera2021design}. While the credo of the deep learning community is to build large neural networks that can solve arbitrary tasks given large enough training examples and compute power, this is not the only way towards "intelligent" algorithms. An alternative is logic AI \cite{heule2017science}. Here, the idea is to translate arbitrary problems to Boolean satisfiability expressions and solve them with powerful SAT solvers.  In \cite{cervera2021design}, the question of designing quantum experiments has been rephrased into logical expressions and solved with \textit{MiniSAT}. It is shown that in some problems, a combination of \textit{Theseus} and the logical approach is faster than the continuous optimization itself. The reason is that the unsatisfiability of candidate solutions is detected quickly with a logical approach, thus guiding the continuous optimizer towards more promising candidates. This approach is in its infancy. Given that the field of logic AI is growing fast due to computational and algorithmic advances, we expect a large increase in interest in this topic.

Deep generative models such as variational autoencoders became a standard tool in fields such as material design \cite{sanchez2018inverse}. Here, an encoder network transforms a (potentially discrete) representation into a continuous latent space. The decoder network is trained to take a point in the latent space and translate it back to the discrete structure. The encoder and decoder together are trained to perform an identity transformation, which by itself is not that interesting. However, as an exciting side effect, the system builds up an internal, continuous latent space that can be shaped during the training and used for gradient-based optimization. For the first time, such a system was demonstrated for quantum optics in \cite{flam2021learning}. The work focuses on understanding \textit{what} the neural networks have learned and how they store the information in their internal latent space. The structure of the latent space shows surprising discrete structures that were then identified with concrete properties of the experimental setups. It will be interesting to see more advanced ways to investigate, navigate and understand the high-dimensional internal representations of neural networks that are built autonomously during training.

 A conceptually related task is the design of superconducting circuits. The quantum behavior of superconducting circuits is defined by a network of inductances, capacitances, and Josephson junctions. As with quantum optical experiments, those systems are conventionally designed by experienced human researchers who aim to find suitable configurations for complex quantum transformations, such as coupling between two well-defined qubits in quantum computers. The search space of possible structures grows exponentially with the number of elements, and thus it quickly becomes infeasible for humans to find solutions for complex tasks. In \cite{menke2021automated}, the authors addressed the question of designing superconducting circuits for the first time with a fully automated closed-loop optimization approach and designed a 4-local coupler by which four superconducting flux qubits interact. The algorithm \textit{SCILLA} starts with a discrete circuit topology. The best candidates are further parametrically optimized, either with a direct gradient-based optimization or with an evolutionary approach (to avoid local minima). The final design outperforms the only other (hand-crafted) 4-local coupler in terms of noise resilience and coupling strength.

\textbf{Discovering Quantum Protocols and Discrete Feedback Strategies} --

Discrete building blocks occur not only naturally in the construction of experiments, but also as part of discrete temporal sequences, specifically sequences of quantum gates and other operations. These sequences can represent quantum protocols or higher-level control strategies for quantum devices. The hardware-level control discussed in the previous section \ref{sec:hardware-level-control} could then be considered as a tool to implement the individual building blocks (e.g. an individual gate). In the following we will deal with protocols that also contain elements of feedback or other actions that are not merely unitary gates. The task of quantum circuit synthesis (building up unitaries out of elements) will be discussed further below. 

\begin{figure}
    \centering
    \includegraphics[width=\columnwidth]{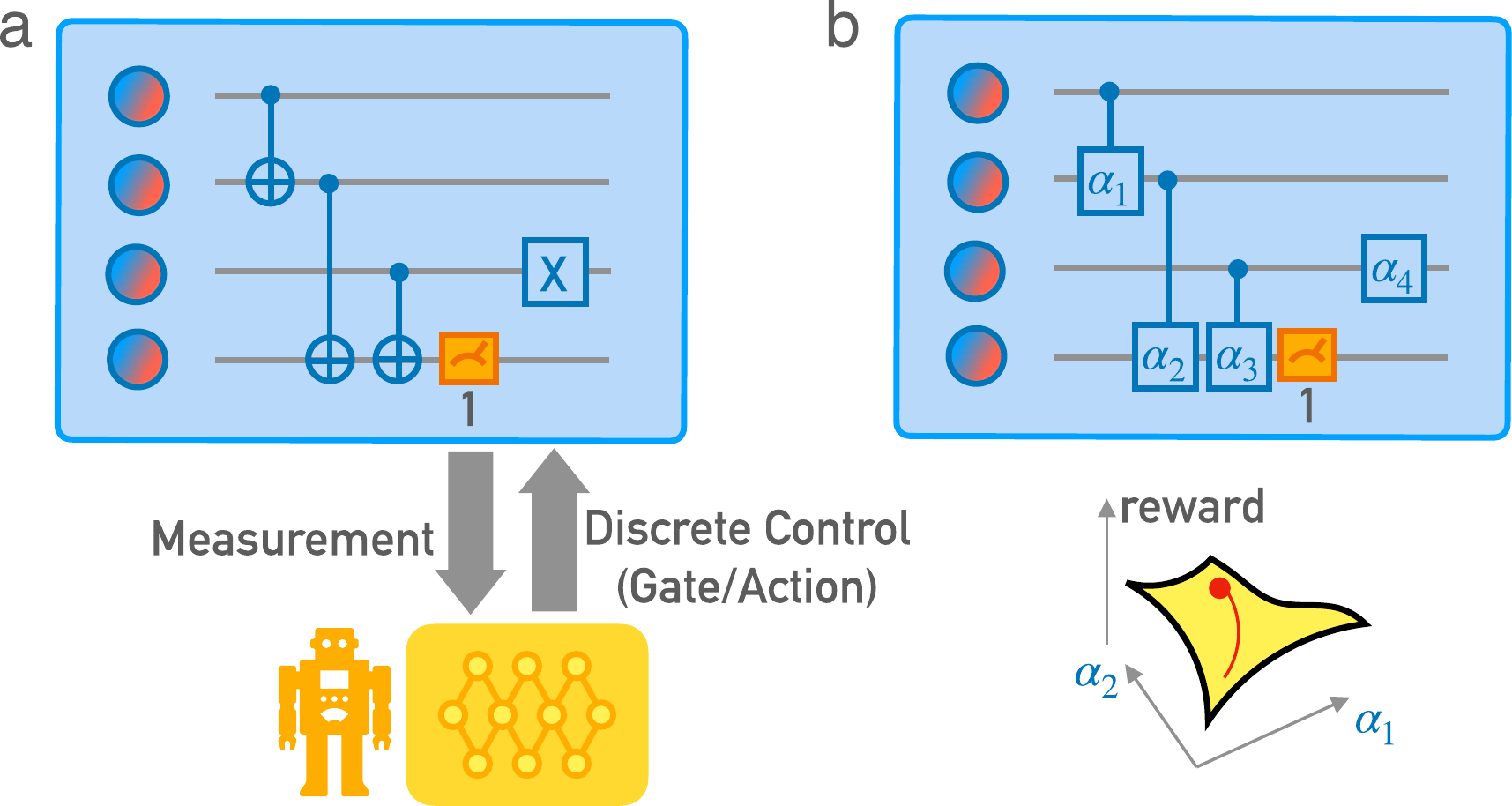}
    \caption{Discovery of Quantum Circuits and Feedback Strategies with Discrete Gates. (a) A reinforcement-learning agent acts on a multi-qubit system by selecting gates, potentially conditioned on measurement outcomes, finding an optimized quantum circuit or quantum feedback strategy. (b) A fixed layout quantum circuit with adjustable parameters that can be optimized via gradient ascent to achieve some goal like state preparation or variational ground state search (possibly including feedback).}
    \label{fig:circuit}
\end{figure}

Reinforcement learning is one suiting tool for the automated discovery of such sequences. This was first analyzed in \cite{fosel_reinforcement_2018}, using deep RL, where the goal was to discover a strategy for quantum error correction in a quantum memory register made of a few qubits. This involves applying discrete unitary gates, which are conditioned on the outcomes of measurements, i.e. 'real-time' feedback executed during the control sequence, Fig.~\ref{fig:circuit}a. Since the aim of \cite{fosel_reinforcement_2018} was primarily to find quantum error correction strategies, we will discuss some more aspects separately in the upcoming section \ref{sec:Quantum_Error_Correction}.

A reinforcement learning technique was subsequently also applied to the rediscovery of implementations for quantum communication protocols \cite{wallnofer_machine_2020}. There, the authors set up the task as an RL problem and explain the similarity of quantum communication and RL with the following intuition: A quantum communication protocol is a sequence of operations that leads to the desired outcome. Similarly, an RL agent learns a policy, that is, to perform sequences of operations that maximize a reward function. The authors task the RL agent to rediscover several important quantum communication schemes such as quantum teleportation, quantum state purification, or entanglement swapping. Each of these tasks can be written as a simple network, where the nodes stand for the involved parties and edges indicate classical or quantum correlations between them. Let us take the quantum teleportation protocol as an example (the others follow similar ideas). The environment is a three-node network (the incoming unknown quantum state A, the sender B, and the receiver C). The environment starts with pre-shared entanglement between B and C. The agent now has to find a correct sequence of local measurements and classical communication steps that teleports the quantum state from A to C. After performing up to 50 operations, the transformations are evaluated, and the agent gets a (binary) reward for whether it succeeded or not. Over 100k trials, the agent finds with high probability an action-efficient strategy to perform the task. As there is no feedback from the environment and a model of the system exists, RL agents are not the only option to solve these questions, and direct gradient-based methods can be used for discovering quantum protocols.

\textbf{Quantum Circuits} -- The design of quantum circuits has some relation to the design of quantum experiments. In both situations, a discrete set of parametrized elements are carefully connected to form the topology of the circuit or experiment, while the continuous parameters of the elements (such as phases) are optimized. In contrast to quantum circuits that can use a universal gate set, for experimental design or the quantum protocols considered above it is not clear in the beginning whether certain targets can be reached with the available resources.

Quantum circuit design problems broadly fall into two branches: First, quantum circuit synthesis (QCS) (sometimes called quantum circuit compilation) addresses the problem of how to build from scratch a circuit that performs a specific task. Second, quantum circuit optimization (QCO) aims at turning a given circuit into a simplified, logically equivalent circuit.

The problem of quantum circuit synthesis is translating an algorithm into elementary gates from a finite universal set. There are two situations -- the fully discrete case, where gates are fixed, and the case where gates can be tuned via continuous parameters (e.g. rotation angles). The first application of machine-learning techniques for the de-novo generation of quantum circuits used genetic algorithms \cite{williams1998automated}. The algorithm had access to  a set of single- and two-qubit gates and was tasked to rediscover the quantum circuit for quantum teleportation. It indeed found the correct circuit with significantly less evaluations than an exhaustive search and could also present different solutions not discussed in the literature before. Genetic algorithms are a powerful tool for discrete discovery. Thus, related approaches are still being used two decades later \cite{mcdonald2013genetic, las2016genetic}.

The problem of designing a quantum circuit from an unparameterized set of elementary gates can be formulated as a reinforcement learning problem. In \cite{zhang2020topological}, the authors introduce a deep reinforcement learning algorithm that translates an arbitrary single-qubit gate into a sequence of elementary gates from a finite universal set for a topological quantum computer. The authors apply their algorithms to Fibonacci anyons and discover high-quality braiding sequences. Similar machine learning techniques have been applied to the quantum circuits for gate-based quantum computers \cite{moro2021quantum}. The design of unparameterized quantum circuits has an important application for fault-tolerant quantum computation, where only a finite list of unparameterized gates can be applied to the logical qubits.

In general, quantum circuits can have tunable parameters -- for instance, parameterized \texttt{X}-gates. In that case, the problem has both a discrete and continuous element. An important application for this task is the hardware-aware design of circuits\cite{cincio2018learning,cincio2021machine}. Here, the algorithms consider the circuit's connectivity on the one hand and noise contributions on the other. The main goal is to find shallow circuits that are more noise-resistant than other (textbook) implementations of circuits. The authors of \cite{cincio2018learning,cincio2021machine} present an algorithm that can significantly outperform textbook solutions for various state generation tasks in terms of fidelity under realistic noise conditions. The design of discrete circuits with continuous parameters has also been approached with deep reinforcement learning \cite{yao2022noise}.

A crucial and  heavily investigated topic in the area of quantum circuit design is the parametric optimization of a constant circuit topology. This task is essential for hybrid quantum-classical variational quantum algorithms (VQA) that can run on near-term quantum computing hardware \cite{cerezo2021variational,bharti2022noisy}, as well as for quantum machine learning, where such parameterized circuits are used as quantum neural networks \cite{dunjko2020non}. We will not go into the details of these topics, but refer the reader to the excellent reviews on this topic.

We will mention only a few selected but important results and ideas for completeness. First, it has been discovered that the gradients in a randomly initialized parametrized quantum circuit vanish exponentially with the number of qubits \cite{mcclean2018barren}, which has become one of the main challenges in the field of quantum machine learning and VQA, Fig.~\ref{fig:circuit}b. An important question then is how an expressive initial state  of the circuit (ansatz) allows for the efficient machine-learning-based optimization of the circuits \cite{sim2019expressibility}. Some exciting approaches involve reinforcement learning that explores economic and expressive initial ans\"{a}tze \cite{ostaszewski2021reinforcement} or ideas that are inspired from neural network architecture search \cite{zhang2021neural}. Besides the direct gradient-based optimization of parametrized quantum circuits, different approaches try to avoid the problem of vanishing gradients, by employing reinforcement learning \cite{yao2021reinforcement}, using ML-based prediction of suitable initial parameters (rather than optimizing the parameters directly) \cite{verdon2019learning} or advanced gradient-free approaches that are naturally not susceptible to the barren plateau problem \cite{anand2021natural}. 

An interesting recent application of VQA-based systems is the \textit{quantum-computer-aided design} of quantum hardware\cite{kyaw2021quantum,kottmann2021quantum,liu2021quantum}. As described at the beginning of this chapter, the AI-based design of new quantum hardware on a classical computer has the problem of  memory requirements  increasing exponentially with the system size. One way to overcome this problem is to outsource the computation of the expensive quantum system to a quantum computer. Here, the problem of designing new multi-qubit couplers for superconducting quantum computers or the design of new quantum optics hardware can be rephrased as in a VQA-style problem. A classical AI algorithm changes the parameter of a parameterized quantum circuit to minimize a fidelity function computed from the outcome of the quantum computation. After convergence, a mapping translates the final parametrized quantum circuit into the specific quantum hardware. This approach has been experimentally demonstrated in a proof-of-principle three-qubit superconducting circuit \cite{liu2021quantum}.

In general, it is not guaranteed that a direct compilation of an algorithm already yields the most efficient implementation of a quantum circuit. A powerful classical method to simplify (compile) quantum circuits is the ZX formalism \cite{duncan2020graph}, which reformulates the circuit into a graph, where predefined rules identify simplifications. However, this and similar approaches have been formulated in a hardware-independent way, operating on a global level. Alternatively, this problem can be approached by RL algorithms \cite{fosel_quantum_2021} that can autonomously simplify circuits, for example, in terms of circuit depth or gate counts, and this enables easily taking into account concrete hard-ware constraints. In \cite{fosel_quantum_2021} this approach was developed and found superior to simulated annealing (tested for circuits of up to 50 qubits). It has the potential to become an important tool for simplifying quantum circuits in the future. 

\textbf{Future Challenges and Opportunities} --
The simulation of quantum experiments becomes expensive as soon as the system grows in size. Neural networks could autonomously find approximate predictions for the dynamics of the quantum system. Such supervised systems need a lot of training examples. Thus the trade-off between the creation of training examples and the computational benefit of an approximation needs to be investigated.

The design of new experiments or hardware can not only be seen as \textit{optimization} (in the sense of making an existing structure better) but as \textit{discovery} in which we create new ideas that did not exist before, as shown in \cite{krenn2021conceptual}. This point of view shows how machines can creatively contribute to science and act as an inspiration for human scientists. There will be a great potential for expanding these ideas. Automatic extraction of understandable building blocks ('subroutines') can help with this challenge.

\subsection{Quantum Error Correction}
\label{sec:Quantum_Error_Correction}

The ability to correct errors in a quantum computing device will be indispensable to realizing beneficial applications of quantum computation since real-world devices are not coherent enough to run an error-free calculation. The basic conceptual ideas in this domain are known since the pioneering work of Shor \cite{shor1995scheme} and subsequent developments, most notably the surface code. In any case, the idea is to encode logical qubit information in many physical qubits robustly and redundantly. The presence of errors (like qubit dephasing and decay) must be detected via measurement of so-called syndromes, i.e.~suitably chosen observables (often multi-qubit operators). Finally, a good way to interpret the observed syndromes and apply some error correction procedure must be found. Despite the knowledge of good encodings and suitable syndromes, it remains a challenging problem how to best implement those in practice, for a given quantum device, with its available gate set and topology of connections between the qubits, and how to optimize them for a given noise model.

It has been recognized early on that machine learning methods could potentially be of great help in this domain. The tasks can naturally be divided into three categories.

\begin{figure}
    \centering
    \includegraphics[width=\columnwidth]{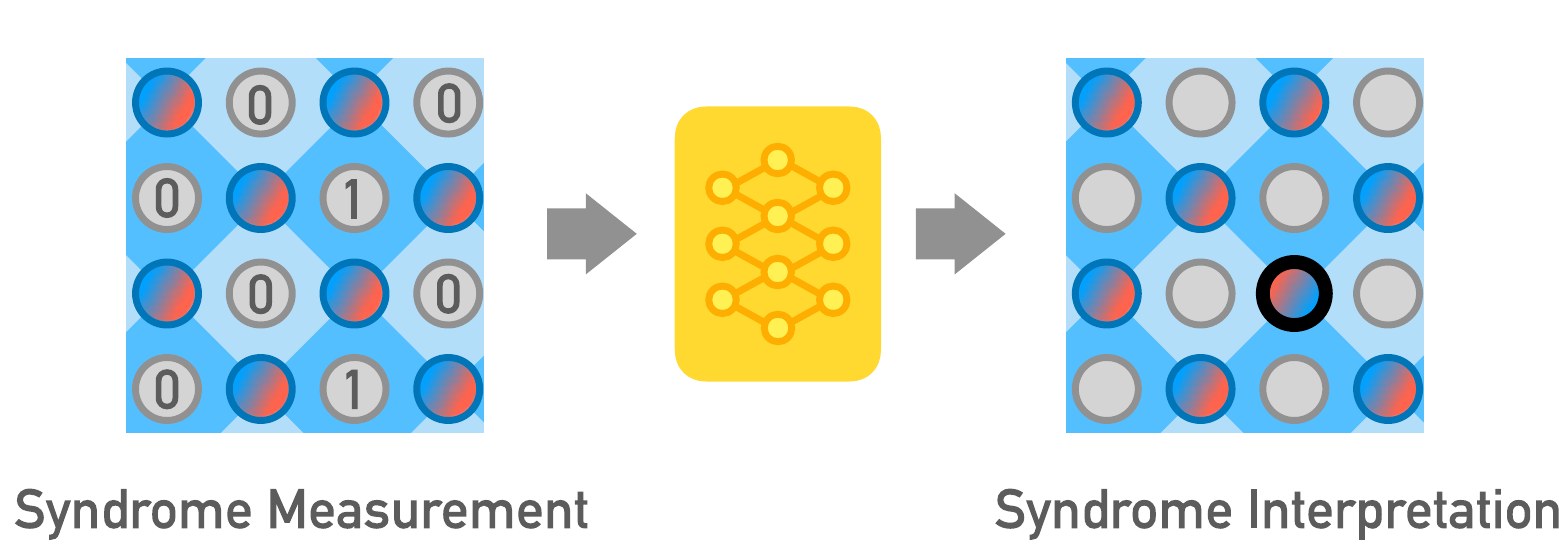}
    \caption{Quantum Error Correction. Syndrome interpretation in a surface code as a task that a neural network can be trained to perform.}
    \label{fig:QEC}
\end{figure}

{\bf Syndrome interpretation} -- On the simplest level, we already assume an existing encoding and a fixed set of syndromes. The task then is to find the optimal way to interpret the observed syndrome, e.g.~deciding which qubits are likely erroneous and must be corrected, see Fig.~\ref{fig:QEC}. This can be phrased as a supervised learning problem, where some errors are simulated, the syndrome is fed into a network, and the network must announce the location of the errors. In practice, the surface code is the most promising QEC architecture, and deducing the error from the syndrome is not trivial, though non-ML algorithms exist. Multiple works therefore trained neural networks to yield ``neural decoders''  \cite{torlai_neural_2017,krastanov_deep_2017,varsamopoulos2017decoding,baireuther2018machine}. In one early example \cite{torlai_neural_2017}, a modified, restricted Boltzmann machine was used, with two types of visible units, corresponding to syndrome and underlying error configuration. This was then trained on a data set of such pairs. Afterward, the machine could be used to sample the errors compatible with an observed syndrome. It is also possible to use reinforcement learning to discover better strategies in more complicated situations. In  \cite{sweke2020reinforcement}, this was applied to the surface code, exemplified in a situation with faulty syndrome measurements. 

{\bf Code search} --
Going one step further, the question arises whether a machine can also find better codes. It is helpful to take an existing code and modify it. The surface code is usually formulated on a simple square lattice, but it can be implemented on more complicated geometries, both periodic and even more generally a-periodic. In \cite{nautrup_optimizing_2019}, a reinforcement learning agent was asked to optimize the connectivity of a surface code given a number of data qubits. It was able to find the best performing code in scenarios important for real experiments. These included biased noise (not all error channels equally strong) or spatially localized noise (higher error rates in the vicinity of some qubits). The agent found interesting nontrivial connectivities as optimal solutions. Due to the availability of highly efficient simulation tools for establishing the performance of surface codes, the authors of \cite{nautrup_optimizing_2019} were able to go up to 70 data qubits.

Autonomous quantum error correction consists of an experimental configuration that can intrinsically correct certain types of errors without active feedback. This idea can be implemented by introducing carefully additional drives and dissipation. The discovery of such mechanisms in real physical systems, under strict experimental constraints, is highly nontrivial. The authors of \cite{wang2022automated} show the automated discovery of an autonomous QEC that could be applied to Bosonic systems. The goal is to find an encoding of a logical qubit that is robust under the dynamics of the system. The algorithm denoted \textit{AutoQEC} can then discover such an encoding by maximizing the average fidelity of the logical qubit. \textit{AutoQEC} is further constrained to consider only systems within experimental capabilities. Indeed, the authors discover a new quantum code, denoted $\sqrt{3}$-code, that has a longer lifetime than previously studied systems with the same concrete experimental constraints. The authors go on and, inspired by their numerical discovery, derive the analytical, logical state and analyze the new autonomous QEC system further.

{\bf Full QEC protocol discovery} -- Finally, one can adopt the attitude that neither the code itself, the code family, or any other ingredients are assumed. In that case, one starts from scratch, and the goal of the machine is to discover ways to preserve the quantum information with high fidelity for as long as possible. In other words, it (re-)discovers all aspects of QEC and error mitigation, adapted to the given platform and noise model. Such an ab-initio approach was demonstrated in \cite{fosel_reinforcement_2018}, which we already mentioned above in the context of RL for feedback. Given a few qubits with arbitrary connectivity and gate set, as well as an arbitrary noise model, the agent is asked to preserve the quantum information as long as possible. To solve this challenging task, additional generally applicable insights were required, e.g. introducing a reward that can measure the amount of surviving quantum information without having discovered a proper decoding sequence. Beyond approaches that fall in the family of stabilizer codes, the same agent also discovered noise mitigation techniques based on adaptive measurements. The advantage of such an ab-initio approach is its flexibility, but the price to pay is that so far, it only works for a handful of qubits due to the effort required in simulations. A future challenge would be finding ways to make the RL work directly on the experiment.

\textbf{Future Challenges and Opportunities} --
Regarding syndrome measurements, an important challenge for the future is ensuring that the neural networks interpreting those measurement results can be deployed in an actual device at sufficient speed: even for the classical algorithms, this is nontrivial. Another challenge is that ab-initio discovery of quantum error correction strategies still relies on simulations, whose numerical effort scales exponentially in the number of qubits (for general quantum dynamics). An interesting direction could also be the co-discovery of autonomous QEC experiments together with QEC feedback strategies in these systems.

\section{Outlook}
\label{sec:outlook}

With all these promising ideas in mind, let us look forward to the year 2035: how do we imagine machine learning to contribute to quantum technologies by that time?

\textbf{Fully controlled and error-corrected quantum systems} -- 
As quantum platforms scale towards ever greater numbers of components and connections, machine learning will provide a way to harness this complexity -- by automatically calibrating and fine-tuning the resulting huge number of parameters adaptively, by discovering optimized experimental setups in the first place, by extracting the maximum amount of information from rather indirect measurements of the underlying phenomena via complex observables, and by finding smart control strategies. Errors in such systems could be corrected by fully automated quantum error correction schemes discovered by an AI. When we think of quantum computers with 1000s of qubits or quantum simulators with even more degrees of freedom, all of this will be a crucial part of the community's toolbox.

\textbf{Specifying goals, not algorithms} -- One important aspect here will be that instead of defining an algorithm that tells the computer {\em how} to achieve some goal, we will typically {\em define the goal itself}. Such goals could be to retain a large fidelity during quantum operations, produce highly entangled states, or have a strong sensitivity to some external signal. The details of how to reach this goal will be left for the computer to discover. This change of perspective will enable a much higher level of description, which is one way to keep ahead of the growing complexity. Ultimately, one might expect that the machine has access to the scientific literature and suggests goals and new experiments autonomously, as demonstrated in material science\cite{tshitoyan2019unsupervised}.

\textbf{Discovering new Algorithms} -- Rather than discovering experiments or feedback strategies, it will be very interesting to see whether ML agents can autonomously discover other higher-logic quantum programs such as quantum algorithms. This task is recently been tackled by large language models for classical algorithms \cite{d2022deep,deepmind2022clrs}, and we expect that similarly quantum algorithms can be discovered with classical machine learning models.

\textbf{How can the human learn?} -- Suppose that computers  will be able to help us find solutions for many of the lower-level and even some higher-level, more conceptual tasks in the domain of quantum technologies. That raises the following notorious question, pervasive throughout machine learning and artificial intelligence: How can we human scientists understand what the machine has learned? Do we need to open the black box of neural networks, or can we use the algorithms as a source of inspiration in a different way \cite{krenn2022scientific}? We argue that while improved performance in the task at hand is great, being able to understand the essence of what the machine has discovered is crucial for the result to become of much wider applicability. In general, gaining \textit{understanding} has been called the \textit{essential aim of science}\cite{de2017understanding}. Here approaches where the solution involves discrete steps (e.g. discrete actions of an agent) or logic-based AI seem to be easier to interpret, explain and understand than results from deep-learning-based methods. The field of symbolic regression (which extracts discrete explanations of neural network predictions) might be very fruitful in this approach.

\textbf{What needs to be done?} --
To attain the visions described above, our community may adopt some proven methodologies from other areas. The idea of fair benchmarks and competitions is one of the powerful driving forces in the development of ML algorithms. One of the most famous examples being the ImageNet data set which provided the basis for a revolution of ML-based computer vision systems \cite{deng2009imagenet}. This idea was adapted in other fields of science that apply AI methodologies, such as material discovery\cite{brown2019guacamol,polykovskiy2020molecular}. In contrast, the field of AI in quantum technology, at the moment, appears more like the wild west. There are no clear ways how to compare approaches from different papers, because most works apply their approaches to slightly different tasks, making them incomparable. We believe that fair and suitably curated benchmark data sets will steer the development of powerful and ever more generally applicable AI algorithms in quantum technology. The data sets could consist of simulated or (in the best case) experimental data for data interpretation tasks. Likewise, to facilitate the discovery of experimental setups and protocols, the community can develop a selection of well-curated objective functions and a set of simulated environments describing important prototypical quantum devices (see \texttt{SciGym} for a first attempt at this \cite{scigym}). In a similar direction, we expect that cloud access to real quantum experiments will become available for significantly more systems. AI algorithms can then be trained on the data from these real machines with specific experimental constraints (such as connectivity or noise). This will boost the capabilities of algorithms that deal with important, real-world systems.

Finally, what Alan Turing remarked in his visionary article on intelligence and learning machines \cite{turing1950computing} is also valid here, in the field of machine learning applied to quantum technologies: "We can only see a short distance ahead, but we can see plenty there that needs to be done."

\section{Acknowledgments}
The authors thank S\"oren Arlt and Xuemei Gu for helpful comments on the manuscript. F.M. acknowledges funding by the Munich Quantum Valley, which is supported by the Bavarian state government with funds from the Hightech Agenda Bayern Plus. 

\bibliography{review}

\end{document}